\documentclass{sig-alternate-05-2015}
\usepackage[numbers,sort&compress]{natbib}
\usepackage{balance}
\usepackage{color, colortbl, fancyvrb,alltt}
\definecolor{LightBlue}{rgb}{0.749,0.75,1.0}
\definecolor{LightOrange}{rgb}{0.99,0.799,0.6} 
\newcommand{\myarr}{$\rightarrow$} 

\newcommand{\THESYSTEM}{\textsf{PrivInfer}\xspace}
\usepackage{xspace}
\usepackage{mathpartir}
\usepackage{stmaryrd}
\usepackage{color}
\usepackage{float}
\usepackage{listings}
\usepackage{xfrac}
\usepackage{bbm}
\usepackage{bbold}
\usepackage{url}
\usepackage{paralist}
\usepackage{bussproofs}
\usepackage{xifthen}
\usepackage{graphicx}
\usepackage{subcaption}

\pdfoutput=1

\DeclareCaptionFormat{myformat}{#1#2#3\hrulefill}
\usepackage{amsthm}
\usepackage{mathabx}

\theoremstyle{plain}
\newtheorem{lemma}{Lemma}[section]
\newtheorem{theorem}{Theorem}[section]

\theoremstyle{definition}
\newtheorem{definition}{Definition}[section]

\theoremstyle{corollary}
\newtheorem{corollary}{Corollary}[section]

\renewcommand{\implies}{\Rightarrow}

\usepackage{listings}

\usepackage[T1]{fontenc}
\usepackage{microtype}

\usepackage[scaled]{beramono}
\newcommand\Small{\fontsize{8.2pt}{8.4pt}\selectfont}
\newcommand*\LSTfont{\Small\ttfamily\SetTracking{encoding=*}{-60}\lsstyle}

\definecolor{DarkGreen}{rgb}{0.1,0.5,0.1}
\definecolor{DarkRed}{rgb}{0.5,0.1,0.1}
\definecolor{DarkBlue}{rgb}{0.1,0.1,0.5}
\usepackage{hyperref}
\hypersetup{
    unicode=false,          
    pdftoolbar=true,        
    pdfmenubar=true,        
    pdffitwindow=false,      
    pdftitle={},    
    pdfauthor={}
    pdfsubject={},   
    pdfnewwindow=true,      
    pdfkeywords={keywords}, 
    colorlinks=true,       
    linkcolor=DarkRed,          
    citecolor=DarkGreen,        
    filecolor=DarkRed,      
    urlcolor=DarkBlue,          
}
\usepackage[capitalise]{cleveref}

\crefname{section}{\S}{\S}
\Crefname{section}{\S}{\S}

\newcommand{\ra}{\rightarrow}

\def\reals{\ensuremath{\mathbb{R}}\xspace}

\def\vars{\ensuremath{\mathcal{X}}\xspace}

\def\pcf{\ensuremath{\mathbf{PCF}}\xspace}
\def\pcfp{\ensuremath{\mathbf{PCF}_{p}}\xspace}

\def\ctts{\ensuremath{\mathcal{C}}\xspace}

\newcommand{\coty}[1]{\widetilde{#1}}

\def\kwcase{\mathtt{case}}
\def\kwwith{\mathtt{with}}

\def\kwletrec{\mathtt{letrec}}
\def\kwin{\mathtt{in}}
\def\kwunit{\mathtt{return}}
\def\kwbind{\mathtt{mlet}}

\def\kwlist{\mathtt{list}}

\newcommand{\slam}[2]{\lambda {#1} .\, {#2}}

\newcommand{\sletrec}[4]{\kwletrec^{#1}\ {#2}\ {#3} = {#4}}

\newcommand{\scase}[2]{\kwcase\ {#1}\ \kwwith\ {#2}}
\newcommand{\sbranch}[2]{{#1} \Rightarrow {#2}}
\newcommand{\sunitM}[1]{\kwunit\ {#1}}
\newcommand{\sbindM}[3]{\kwbind\ {#1} = {#2}\ \kwin\ {#3}}

\newcommand{\stfun}[2]{{#1} \rightarrow {#2}}

\newcommand{\stmod}[1]{\mathfrak{M}[{#1}]}

\newcommand{\stunit}[0]{\bullet}
\newcommand{\stbool}[0]{\mathbb{B}}
\newcommand{\stnat}[0]{\mathbb{N}}

\newcommand{\streal}[0]{\mathbb{R}}

\newcommand{\stlist}[1]{{#1}\ \kwlist}

\def\lvmark{\triangleleft}
\def\rvmark{\triangleright}

\renewcommand{\l}[1]{#1_\lvmark}
\renewcommand{\r}[1]{#1_\rvmark}
\def\svar{\mathfrak{s}}

\newcommand{\rembed}[1]{|{#1}|}
\newcommand{\rrembed}[1]{\|{#1}\|}
\newcommand{\rmark}[1]{{#1}^{\Join}}

\def\pvars{\ensuremath{\mathcal{X}_{\mathcal{P}}}\xspace}
\def\rvars{\ensuremath{\mathcal{X}_{\mathcal{R}}}\xspace}

\newcommand{\expr}[0]{\ensuremath{\mathcal{E}}}
\newcommand{\rexpr}[0]{\ensuremath{\expr^{\Join}}}

\newcommand{\rtypes}[0]{\ensuremath{\mathcal{T}}\xspace}
\newcommand{\rassert}[0]{\ensuremath{\mathcal{A}}\xspace}

\newcommand{\rtmod}[3]{\mathfrak{M}_{#1,#2}[{#3}]}

\newcommand{\rtprod}[3]{\Pi ({#1} :: {#2}) .\, {#3}}
\newcommand{\rtref}[3]{\{ {#1} :: {#2} \vbar {#3} \}}

\def\rfalse{\bot}
\def\rtrue{\top}

\newcommand{\fquant}[4]{{#1}\ ({#2} : {#3}) .\, {#4}}
\newcommand{\rquant}[4]{{#1}\ ({#2} :: {#3}) .\, {#4}}

\def\quantvar{\mathcal{Q}}
\def\rformc{\mathcal{C}}

\newcommand{\rmapx}[4]{%
{#1}{\left\{\substack{\l{#2} \,\mapsto\, {#3}\\ \r{#2} \,\mapsto\, {#4}}\right\}}}

\newcommand{\renv}[1]{\mathcal{#1}}

\newcommand{\tyinterp}[1]{\llbracket {#1} \rrbracket}
\newcommand{\interp}[2]{\llbracket {#2} \rrbracket_{#1}}
\newcommand{\rinterp}[2]{\llparenthesis {#2} \rrparenthesis_{#1}}

\def\tyle{\preceq}

\DeclareMathOperator{\dom}{dom}

\newcommand{\vbar}[0]{\mathrel{|}}
\newcommand{\vv}[1]{\overline{#1}}
\newcommand{\vvi}[2]{[{#2}]_{#1}}

\newcommand{\tmod}[1]{\mathfrak{M}[{#1}]}

\newcommand{\lift}[2]{\mathcal{L}_{#1}(#2)}

\def\R{\mathbb{R}}

\newcommand{\rplusinfty}{\ensuremath{\overline{\R}^+}}

\newcommand{\rplus}{\R^+}

\newcommand{\M}[2][]{%
            \ifthenelse{\isempty{#1}}%
              {\mathsf{M}\, {#2}}
              {\mathsf{M}_{#1}\, {#2}}}

\newcommand{\real}[0]{\ensuremath{\mathbb{R}}\xspace}

\renewcommand{\P}{\mathcal{P}}

\newcommand{\observe}[3]{\mathsf{observe}\, #1\Rightarrow #2\,
  \mathsf{in}\, #3}
\newcommand{\binfer}[1]{\mathsf{infer}(#1) }
\newcommand{\bran}[1]{\mathsf{ran}(#1) }
\newcommand{\bernoulli}[1]{\mathsf{bernoulli}(#1) }
\newcommand{\normal}[2]{\mathsf{normal}(#1,#2) }

\newcommand{\laplace}[2]{\mathsf{lapMech}(#1,#2) }
\newcommand{\gauss}[2]{\mathsf{gaussMech}(#1,#2) }
\newcommand{\betad}[2]{\mathsf{beta}(#1,#2) }
\newcommand{\uniform}{\mathsf{uniform}() }
\newcommand{\stdist}[1]{\mathfrak{D}[{#1}]}
\newcommand{\exponential}[3]{\mathsf{expMech}(#1,#2,#3) }
\newcommand{\fdiv}{\ensuremath{f}}
\newcommand{\distr}{\mathcal{D}}

\newcommand{\charfun}{\ensuremath{\mathbb{1}}}

\begin{document}
\CopyrightYear{2016}
\setcopyright{acmlicensed}
\conferenceinfo{CCS'16,}{October 24 - 28, 2016, Vienna, Austria}
\isbn{978-1-4503-4139-4/16/10}\acmPrice{\$15.00}
\doi{http://dx.doi.org/10.1145/2976749.2978371}

\clubpenalty=10000 
\widowpenalty = 10000
\title{Differentially Private Bayesian Programming}
\numberofauthors{7}
\author{
  \alignauthor
  Gilles Barthe \\
  \affaddr{IMDEA Software}
  \alignauthor
  Gian Pietro Farina\titlenote{
    Partially supported by NSF grants CNS-1237235, CNS1565365 and by
    EPSRC grant EP/M022358/1.}\\
  \affaddr{University at Buffalo, SUNY}
  \and
  \alignauthor
  Marco Gaboardi{\raisebox{12pt}{$\scriptstyle *$}}\\
  \affaddr{University at Buffalo, SUNY}
  \alignauthor
  \mbox{Emilio Jes\'us Gallego Arias}  \\
  \affaddr{CRI Mines-ParisTech}
  \alignauthor
  Andy Gordon\\
  \affaddr{Microsoft Research}
  \and
  \alignauthor
  Justin Hsu\titlenote{%
    Partially supported by NSF grants $\#1065060$ and $\#1513694$, and a grant
    from the Simons Foundation ($\#360368$ to Justin Hsu).}\\
  \affaddr{University of Pennsylvania}
  \alignauthor
  Pierre-Yves Strub\\
  \affaddr{IMDEA Software}
}
\maketitle
\begin{abstract}
  We present \THESYSTEM, an expressive framework for writing and
  verifying differentially private Bayesian machine learning algorithms. Programs
  in \THESYSTEM are written in a rich functional probabilistic programming language
  with constructs for performing Bayesian inference. Then, differential
  privacy of programs is established using a relational refinement type
  system, in which refinements on probability types are indexed by a
  metric on distributions. Our framework leverages recent developments
  in Bayesian inference, probabilistic programming languages, and in
  relational refinement types. We demonstrate the expressiveness
  of \THESYSTEM by verifying privacy for several examples of private
  Bayesian inference.
\end{abstract}

\section{Introduction}
Differential privacy~\citep{DMNS06} is emerging as a gold standard in
data privacy. Its statistical guarantee ensures that the probability
distribution on outputs of a data analysis is almost the
same as the distribution on outputs from a hypothetical dataset that differs in one individual. A
standard way to ensure differential privacy is by
perturbing the data analysis adding some statistical noise. The
magnitude and the shape of noise must provide a protection to the influence of an
individual on the result of the analysis, while ensuring that
the algorithm provides useful results. Two properties of differential privacy are especially relevant for this work: (1)
\emph{composability}, (2) the fact that differential privacy works well on large
datasets, where the presence or absence of an individual has limited impact.
These two properties have led to the design of tools for
differentially private data analysis. Many of these tools use
programming language techniques to ensure that the resulting programs
are indeed differentially private~\citep{PINQ09,ReedPierce10,POPL:BKOZ12,GaboardiHHNP13,EignerM13,DBLP:conf/csfw/BartheDGKB13,BartheO13,BartheGAHRS15,conf/popl/EbadiSS15}. Moreover, property (2) has encouraged the
interaction of the differential privacy community with the machine
learning community to design privacy-preserving machine learning
techniques, e.g.~\citep{DworkRV10,Chaudhuri:2011,HardtLM12,Zhang:2014:PrivBayes}.
At the same time, researchers in \emph{probabilistic programming} are exploring
programming languages as tools for machine learning. For example, in Bayesian
inference, probabilistic programming allows data analysts to represent the
probabilistic model, its parameters,
and the data observations as a specially crafted program. Given this
program as input, we can then use inference algorithms to produce a distribution
over the parameters of the model representing our updated beliefs on them.
Several works have explored the design of programming languages to
compute efficiently the updated beliefs in order to produce efficient
and usable tools for machine learning,
e.g.~\citep{LunnTBS00,Pfeffer01,DBLP:conf/uai/GoodmanMRBT08,InferNET12,GordonGRRBG14,TolpinMW15}. 

Recently, research in Bayesian inference and machine learning has turned to
privacy-preserving Bayesian inference
\citep{DBLP:conf/nips/WilliamsM10,DimitrakakisNMR14,Zheng16,ZhangRD16},
where the observed data is private.  Bayesian inference is a deterministic
process, and directly releasing the posterior distribution would violate differential
privacy.  Hence, researchers have developed techniques to make Bayesian
inference differentially private.  Basic techniques add noise on the input data,
or add noise on the result of the data analysis, while more advanced techniques
can ensure differential privacy by taking samples from the posterior
distribution instead of releasing the posterior distribution explicitly. The
diversity of approaches makes Bayesian inference an attractive target for
verification tools for differential privacy.

In this work we present \THESYSTEM, a programming framework combining
verification techniques for differential privacy with learning techniques for
Bayesian inference in a functional setting. \THESYSTEM consists of two main
components: a probabilistic functional language for Bayesian inference, and a
relational higher-order type system that can verify differential privacy for
programs written in this language.  The core idea of Bayesian learning is to use
conditional distributions to represent the beliefs updated after some
observations. \THESYSTEM, similarly to other programming languages for inference
models conditioning on data explicitly. In particular, we extend the functional
language \pcf with an \textbf{observe} statement.

Even though the output of Bayesian inference output is a probability
distribution, it is still a deterministic process. To guarantee differential
privacy, we must inject some randomness into the inference process.  To handle
these two views on distributions, \THESYSTEM distinguishes between
\emph{symbolic} distributions and \emph{actual} distributions. The former
represent the result of an inference, while the latter are used to represent
random computations, e.g.  differentially private computations
(\emph{mechanisms}). We parametrize our language with an algorithm to perform
Bayesian inference returning symbolic distributions, and mechanisms to ensure
differential privacy returning actual distributions. 

Differential privacy is a probabilistic 2-property, i.e. a property expressed
over pairs of execution traces of the program. To address this challenge, we
use an approach based on approximate relational higher-order refinement type
system called $\mathsf{HOARe}^2$ ~\citep{BartheGAHRS15}. We show how to extend
this approach to deal with the constructions that are needed for Bayesian
inference like the $\mathsf{observe}$ construct and the distinction between
symbolic and actual distribution. 

Another important aspect of the verification of differential privacy is
reasoning about the \emph{sensitivity} of a data analysis. This measures the
influence that two databases differing in one individual can have on the output.
Calibrating noise to sensitivity ensures that the data analysis provides
sufficient privacy. In Bayesian inference, the output of the computation is a
distribution (often defined by a few numeric parameters) for which one can
consider different measures. A simple approach is to  considered standard
metrics (Euclidean, Hamming, etc.) to measure the distance between the
parameters. Another more approach is to consider distances
between distributions, rather than the parameters. 

The  type system of \THESYSTEM allows one to reason about the parameters of a
distribution, using standard metrics, but also about the distribution itself
using \emph{\fdiv-divergences}, a class of probability metrics including some
well known examples like total variation distance, Hellinger distance, KL
divergence, etc.  In summary, we extend the approach of  \THESYSTEM in three
directions:
\begin{itemize}
\item  we provide a relational typing rule for $\mathsf{observe}$ and for $\mathsf{infer}$, 
\item we provide a generalization of the relational type system of $\mathsf{HOARe}^2$ to reason about symbolic
and actual distributions, 
\item we generalize the probability polymonad of $\mathsf{HOARe}^2$ to reason about general
$f$-divergences.
\end{itemize}
The combination of these three contributions allows us to address Bayesian
inference, which is not supported by $\mathsf{HOARe}^2$.

To illustrate the different features of our approach we show how different basic
Bayesian data analysis can be guaranteed differentially private in
three different ways: by adding noise on the input, by adding noise on
the parameters with sensitivity measured using the $\ell_1$-distance
between the parameters, and finally by adding noise on the
distributions with sensitivity measured using \fdiv-divergences.
This shows that \THESYSTEM can be used for a diverse set of Bayesian data analyses.
Summing up, the contributions of our work are:
\begin{itemize}
\item  A probabilistic extension \pcfp of \pcf for Bayesian inference that
  serves as the language underlying our framework \THESYSTEM (\Cref{sec:syntax}). This
  includes an $\mathsf{observe}$ statement as well as primitives for handling symbolic and actual distributions.
\item A higher-order approximate relational type system
  for reasoning about properties of two runs of programs from \THESYSTEM (\Cref{sec:relational-typing}). In particular,
  the type system permits to reason about \fdiv-divergences. The \fdiv-divergences can be used to reason
  about differential privacy as well as about program sensitivity for
  Bayesian inference. The relational type system can also reason about symbolic distributions as well.
\item We show on several examples how \THESYSTEM can be used to reason
  about differential privacy (\Cref{sec:examples}). We will explore three ways to guarantee
  differential privacy: by adding noise on the input, by adding noise
  on the output parameters based on $\ell_p$-norms, and by adding noise on the
  output parameters based on \fdiv-divergences.
\end{itemize}

\section{Bayesian Inference}
\label{sec:motivations}

Our work is motivated by \emph{Bayesian inference}, a statistical
method which takes a \emph{prior} distribution $\Pr(\xi)$ over a parameter $\xi$
and some observed data $x$, and produces the \emph{posterior} distribution
$\Pr(\xi \mid x)$, an updated version of the prior distribution. Bayesian
inference is based on \emph{Bayes' theorem}, which gives a formula for the
posterior distribution:
\begin{equation*} 
  \Pr(\xi \mid x)=\frac{\Pr(x \mid \xi)\cdot \Pr(\xi)}{\Pr(x)}
\end{equation*}
The expression $\Pr(x \mid \xi)$ is the \emph{likelihood}
of $\xi$ when $x$ is observed. This is a function $\mathcal{L}_{x}(\xi)$
of the parameter $\xi$ for fixed data $x$, describing the
probability of observing the data $x$ given a specific value of the parameter $\xi$.
Since the data $x$ is considered fixed, the expression $\Pr(x)$
denotes a normalization constant ensuring that $\Pr(\xi \mid x)$ is a
probability distribution. The choice of the prior reflects the prior knowledge
or belief on the parameter $\xi$ before any observation has been
performed. For convenience, the prior and the likelihood are typically chosen in
practice so that the posterior
belongs to the same family of distributions as the prior. In this case
the prior is said to be \emph{conjugate prior} for the likelihood.
Using conjugate priors, besides being mathematically convenient in the
derivations, ensures that Bayesian inference can be performed by a recursive
process over the data. 

Our goal is to perform Bayesian inference under differential
privacy. We provide the formal definition of differential privacy in
\Cref{def:dp}, but for the purpose of this section it is enough to know that
differential privacy is a statistical guarantee that requires the answer to a
data analysis to be statistically close when run on two
\emph{adjacent} databases, i.e. databases that differ in one
individual. In the \emph{vanilla} version of differential privacy, the notion of ``statistically close''
is measured by a parameter $\epsilon$. 
A typical way to achieve differential privacy is to add random
noise, and we present several primitives for doing this in
\Cref{sec:background}. For one example, the \emph{exponential
mechanism} (denoted $\sf ExpMech_{\epsilon}$) returns a possible
output with
probability proportional to a quality score function ${\sf Q}$.  ${\sf Q}$ takes in input
a database and a potential output for the statistic computed on the database, and
gives each output a
score representing how good that output is for that database.
The privacy and the utility of
the mechanism depend on $\epsilon$ and on the \emph{sensitivity} of
the quality score function,
i.e., how much the quality score can differ for two adjacent databases.

As a motivating example we will consider a simple Bayesian inference task:
learning the bias of a coin from some observations. For example, we can think of
the observations as medical records asserting whether patients from a sample
population have a disease or not. We can perform Bayesian inference to establish
how likely it is to have the disease in the population. We will show how to make
this task differentially private, and verify privacy in \THESYSTEM. 

First, the input of this example is a set of binary observations
describing whether any given patient has the disease. We consider this
the \emph{private} information that we want to protect. We also assume that
the number of patients $n$ is \emph{public} and that the \emph{adjacency} condition
for differential privacy states that two databases
differ in the data of one patient. In our concrete case this means
that two databases $d,d'$ are adjacent if all of their records are
the same except for  one record that is $0$ in one database and $1$
in the other.

While in abstract our problem can be described as estimating the bias of a coin,
we need to be more formal and provide the precise model and the parameters that
we want to estimate. We can incorporate our initial belief on the fairness of
the coin using a prior distribution on the bias $\xi$ given by a \emph{beta
  distribution}. This is a distribution over $[0,1]$ with probability density:
$$
{\sf beta}(\xi \mid a,b)=\frac{\xi^{a-1}(1-\xi^{b-1})}{B(a,b)}
$$
where $a,b\in \rplus$ are parameters and $B$ denotes the beta function. The likelihood is the
probability that a series of i.i.d samples from a Bernoulli distributed random variable with
bias $\xi$ matches the observations.
Using an informal
notation,\footnote{We omit in particular the monadic probabilistic
  constructions. A formal description of this example can be found in
  \Cref{sec:examples}.} we can write the following program in \THESYSTEM:
\begin{align}
{\sf infer}\Big ( {\sf observe}\, \big (\lambda r. {\sf bernoulli}(r)={\sf
  obs}\big )\ {\sf beta}(a,b)\Big )
\label{informal}
\end{align}
The term $\sf infer$ represents an inference
algorithm and the ${\sf observe}$  statement is used to describe the
model. Expression (\ref{informal}) denotes the posterior distribution that is computed using
Bayes' theorem with prior ${\sf beta}(a,b)$ and with
likelihood $(\lambda r. {\sf bernoulli}(r)={\sf obs})$.

Now, we want to ensure differential privacy. We have several options.
A first natural idea is to perturbate the input data using
the exponential mechanism, corresponding to the following program:
$$
{\sf infer}\Big ( {\sf observe} (\lambda r. {\sf bernoulli}(r)={\sf
  ExpMech_\epsilon\, Q\, obs})\ {\sf beta}(a,b)\Big )
$$

The fact that differential privacy is closed under post-processing ensures that
this guarantees differential privacy for the whole program. In more detail, in
the notation above we denoted by ${\sf Q}$ the scoring function. Since ${\sf
  obs}$ is a boolean, we can use a quality score function that gives score $1$
to $b$ if $b={\sf obs}$ and $0$ otherwise. This function has sensitivity $1$ and
so one achieves $(\epsilon,0)$-differential privacy.  This is a very simple
approach, but in some situations it can already be very
useful~\citep{DBLP:conf/nips/WilliamsM10}.

A different way of guaranteeing differential privacy is by adding noise on the
output. In this case the output is the posterior which is a ${\sf
  beta}(a',b')$ for some values $a',b'$. Using again the exponential
mechanism we can consider the following program:
$$
{\sf   ExpMech_\epsilon\, Q}\Big ({\sf infer}\big ( {\sf observe} (\lambda
r. {\sf bernoulli}(r)
={\sf
 obs}) {\sf beta}(a,b)\big )\Big )
$$
In this case, the exponential mechanism is not applied to booleans but instead
to \emph{distributions} of the shape ${\sf  beta}(a',b')$. So, a natural
question is which $\sf Q$ we can use as quality score function and what is its
sensitivity in this case. 

There are two natural choices. The first one is to consider the parameters
$(a',b')$ as a vector and measure the possible distance in term of some metric
on vectors, e.g. the one given by $\ell_1$ norm
$d((a,b),(a',b'))=|a-a'|+|b-b'|$. The second is to consider ${\sf  beta}(a',b')$
as an actual distribution and then use a notion of distance on distributions,
e.g. Hellinger distance $\Delta_{\mathcal{H}}({\sf  beta}(a,b),{\sf
  beta}(a',b'))$.

These two approaches both guarantee privacy, but they have different utility
properties. Our system \THESYSTEM can prove privacy for both approaches.

\section{Background}
\label{sec:background}
\subsection{Probability and Distributions}
In our work we will consider discrete distributions.
Following~\citet{DBLP:journals/fttcs/DworkR14} we will use standard names for
several continuous distributions but we will consider them to be the approximate
discrete versions of these distributions up to arbitrary precision.

We define the set $\distr(A)$ of {\em distributions} over a set $A$ as
the set of functions $\mu:A\to [0,1]$ with discrete
$\mathsf{support}(\mu)=\{x \mid \mu\,x\neq 0 \}$, such that $\sum_{x\in
  A}\mu\,x=1$. 
In our language we will consider only distribution over basic
types, this guarantees that all our distributions are
discrete
(see \Cref{sec:syntax}). 

We will use several basic distributions like $\mathsf{uniform},
\mathsf{bernoulli}$, $\mathsf{normal},\mathsf{beta}$, etc. These are all standard distributions
and we omit their definition here. We will also use some notation to
describe distributions. For instance, given an element $a\in
A$, we will denote by  $\charfun_a$ the
probability distribution that assigns all mass to the value $a$. 
We will also denote by $\mathsf{bind}\, \mu\, M$ the composition of a
distribution $\mu$ over the set $A$ with a function $M$ that takes a value in
$A$ and returns a distribution over the set $B$.


\subsection{Differential Privacy}
\label{sec:dp}

\emph{Differential privacy} is a strong,
quantitative notion of statistical privacy proposed by \citet{DMNS06}. In the
standard setting, we consider a program (sometimes called a \emph{mechanism})
that takes a \emph{private database} $d$ as input, and produces a distribution
over outputs. Intuitively, $d$ represents a collection of data from different
individuals. When two databases $d, d'$ are identical except for a single
individual's record, we say that $d$ and $d'$ are
\emph{adjacent}\footnote{In our concrete examples we will consider
  sometime as \emph{adjacent} also two
  databases that differ by \emph{at most} one individual.}, and we write
$d\, \Phi\, d'$. Then, differential privacy states that the output distributions
from running the program on two adjacent databases should be statistically
similar. More formally:
\begin{definition}[\citet{DMNS06}]
\label{def:dp}
  Let $\epsilon, \delta > 0$ be two numeric parameters, let $D$ be the set of
  databases, and let $R$ be the set of possible outputs. A program $M : D \to
  \distr(R)$ satisfies \emph{$(\epsilon, \delta)$-differential privacy} if
  \[
    \Pr( M(d) \in S ) \leq e^\epsilon \Pr( M(d') \in S ) + \delta
  \]
  for all pairs of adjacent databases $d, d' \in D$ such that $d\ \Phi\ d'$, and
  for every subset of outputs $S \subseteq R$.
\end{definition}
As shown by \citet{POPL:BKOZ12}, we can reformulate differential privacy using a specific statistical
$\epsilon$-distance $\epsilon\textnormal{-}\mathsf{D}$:
\begin{lemma}
\label{lem:eps-dist}
  Let $\epsilon, \delta \in \rplus $. Let $D$ be the set of databases, and let
  $R$ be the set of possible outputs. A program $M : D \to \distr(R)$ satisfies
  \emph{$(\epsilon, \delta)$-differential privacy} iff
  $\epsilon$-$\mathsf{D}(M(d), M(d'))\leq \delta$, where $d,d'$ are adjacent
  databases and
\begin{center}
$\epsilon$-$\mathsf{D}(\mu_1,\mu_2)\equiv\displaystyle \max_{E\subseteq R} \big( \Pr_{x\leftarrow \mu_1} [x\in E] - e^{\epsilon}\cdot\Pr_{x\leftarrow \mu_2}[x\in E] \big)$
\end{center} for $\mu_1, \mu_2\in \distr(R)$.
\end{lemma}
Differential privacy is an unusually robust notion of privacy. It degrades
smoothly when private mechanisms are composed in sequence or in parallel, and it
is preserved under any post-processing that does not depend on the private
database. The following lemmas capture these properties:

\begin{lemma}[Post-processing]
Let $M:D\rightarrow\distr(R)$ be an  $(\epsilon, \delta)$-differentially private program. Let $N:R\rightarrow \distr(R')$ be an
arbitrary randomized program. Then $\lambda d. \mathsf{bind}\, (M\, d)\, N:D\rightarrow\distr(R')$ is $(\epsilon,\delta)$-differentially private.
\end{lemma}
Differential privacy enjoys different composition schemes, we report
here one of the simpler and most used. 
\begin{lemma}[Composition]
Let $M_1:D\rightarrow\distr(R_1)$, and  $M_2:D\rightarrow\distr(R_2)$
respectively $(\epsilon_1,\delta_1)$ and $(\epsilon_2,\delta_2)$
differentially private programs. Let $M:D\rightarrow \distr(R_1\times
R_2)$ the program defined as $M(x)\equiv (M_1(x),M_2(x))$. Then, $M$ is $(\epsilon_1+\epsilon_2,\delta_1+\delta_2)$ differentially private.
\end{lemma}
Accordingly, complex differentially private programs can be easily
assembled from simpler private components, and researchers have proposed a
staggering variety of private algorithms which we cannot hope to summarize here.
(Interested readers can consult \citet{DBLP:journals/fttcs/DworkR14} for a
textbook treatment.)

While these algorithms serve many different purposes, the vast majority are
constructed from just three private operations, which we call \emph{primitives}.
These primitives offer different ways to create private mechanisms from
non-private functions. Crucially, the function must satisfy the following
\emph{sensitivity} property:
\begin{definition}
  Let $k \in \rplus$. Suppose $f : A \to B$ is a function, where $A$ and $B$
  are equipped with distances $d_A$ and $d_B$. Then $f$ is \emph{$k$-sensitive}
  if
  \[
  d_B(f(a), f(a')) \leq k \cdot d_A(a, a')
  \]
  for every $a, a' \in A$.
\end{definition}
Intuitively, $k$-sensitivity bounds the effect of a small change in the input, a
property that is similar in spirit to the differential privacy guarantee.
With this property in hand, we can describe the three basic primitive operations
in differential privacy, named after their noise distributions.

\paragraph*{The Laplace mechanism}
The first primitive is the standard way to construct a private version of a
function that maps databases to numbers. Such functions are also called
\emph{numeric queries}, and are fundamental tools for statistical analysis. For
instance, the function that computes the average age of all the individuals in a
database is a numeric query. When the numeric query has bounded sensitivity, we
can use the \emph{Laplace mechanism} to guarantee differential privacy.

\begin{definition}
  Let $\epsilon \in \rplus$ and let $f : D \to \mathbb{R}$ be a numeric
  query. Then, the \emph{Laplace mechanism} maps a database $d \in D$ to $f(d) +
  \nu$, where $\nu$ is drawn form the Laplace distribution with scale
  ${1/\epsilon}$. This distribution has the following probability density
  function:
  \[
    \text{Lap}_{1/\epsilon}(x) = \frac{\epsilon}{2} \exp( - |x| \epsilon ) .
  \]
  If $f$ is a $k$-sensitive function, then the Laplace mechanism is $(k\epsilon,
  0)$-differentially private.
\end{definition}

\paragraph*{The Gaussian mechanism}

The Gaussian mechanism is an alternative to the Laplace mechanism, adding
Gaussian noise with an appropriate standard deviation to release a numeric
query. Unlike the Laplace mechanism, the Gaussian mechanism does not satisfy
$(\epsilon, 0)$-privacy for any $\epsilon$. However, it satisfies $(\epsilon,
\delta)$-differential privacy for $\delta \in \rplus$.

\begin{definition}
\label{def:gaussMech}
  Let $\epsilon, \delta \in \mathbb{R}$ and let $f : D \to
  \mathbb{R}$ be a numeric query. Then, the \emph{Gaussian mechanism} maps a
  database $d \in D$ to $f(d) + \nu$, where $\nu$ is a drawn from the Gaussian
  distribution with standard deviation
  \[
    \sigma(\epsilon, \delta) = \sqrt{2 \ln(1.25/\delta)}/\epsilon .
  \]
  If $f$ is a $k$-sensitive function for $k < 1/\epsilon$, then the Gaussian
  mechanism is $(k\epsilon, \delta)$-differentially private.
\end{definition}

\paragraph*{The exponential mechanism}

The first two primitives can make numeric queries private, but in many
situations we may want to privately release a non-numeric value. To accomplish
this goal, the typical tool is the \emph{exponential mechanism} \citep{MT07},
our final primitive. This mechanism is parameterized by a set $R$, representing
the range of possible outputs, and a \emph{quality score} function $q : D \times
R \to \mathbb{R}$, assigning a real-valued score to each possible output given a
database.

The exponential mechanism releases an output $r \in R$ with approximately the
largest quality score on the private database. The level of privacy depends on
the sensitivity of $q$ in the database. Formally:

\begin{definition}[\citet{MT07}]
  Let $\epsilon \in \rplus$. Let $R$ be the set of outputs, and $q :
  D \times R \to \mathbb{R}$ be the quality score. Then, the \emph{exponential
    mechanism} on database $d \in D$ releases $r \in R$ with probability
  proportional to
  \[
    \Pr(r) \sim \exp\left( \frac{q(d, r) \epsilon}{2} \right)
  \]
  If $f$ is a $k$-sensitive function in $d$ for any fixed $r \in R$, then the
  exponential mechanism is $(k\epsilon, 0)$-differentially private.
\end{definition}


\subsection{\fdiv-divergences}
As we have seen, differential privacy is closely related to function
sensitivity. To verify differential privacy for the result of probabilistic
inferences, we will need to work with several notions of distance between
distributions. These distances can be neatly described as
\fdiv-\emph{divergences}~\citep{csiszar63}, a rich class of metrics on
probability distributions. Inspired by the definition of relative entropy,
\fdiv-divergences are defined by a convex function \fdiv.  Formally:
\begin{definition}[\citet{csiszarS04}]
\label{def:f-div}
  Let $\fdiv(x)$ be a convex function defined for $x>0$, with
  $\fdiv(1)=0$. Let $\mu_1,\mu_2$ distributions over $A$. Then, the \fdiv-\emph{divergence} of $\mu_1$
  from $\mu_2$, denoted $\Delta_\fdiv(\mu_1\ \mid \mu_2)$ is defined as:
$$
\Delta_\fdiv(\mu_1\ \mid \mu_2)=\sum_{a\in A}\mu_2(a)\fdiv\Big
(\frac{\mu_1(a)}{\mu_2(a)}\Big )
$$  
where we assume $0 \cdot \fdiv(\frac{0}{0})=0$ and
$$0 \cdot \fdiv\Big (\frac{a}{0}\Big )=\lim_{t\to 0}t \cdot \fdiv\Big
(\frac{a}{t}\Big )=a\lim_{u\to \infty}\Big (\frac{\fdiv(u)}{u}\Big ).$$
If $\Delta_\fdiv(\mu_1\ \mid \mu_2)\leq \delta$ we say that $\mu_1$ and
$\mu_2$ are $(\fdiv,\delta)$-close.
\end{definition}

Examples of \fdiv-divergences include \emph{KL-divergence}, \emph{Hellinger}
distance, and \emph{total variation} distance.  Moreover, \citet{BartheO13}
showed how the \emph{$\epsilon$-distance} of Lemma~\ref{lem:eps-dist} can be seen
as an \fdiv-divergence for differential privacy. These \fdiv-divergences are
summarized in \Cref{f_div}.
\begin{table}[t]
  \begin{tabular}{lcl}
    {\bf \fdiv-diverg.}& ${\fdiv}(x)$ & {\bf Simplified form}\\
    \hline\\
    $\mathsf{SD}(x)$&$\frac{1}{2}\ |x-1|$& $\displaystyle{\sum_{a\in
      A}}\frac{1}{2}|\mu_1(a)-\mu_2(a)|$\\
    $\mathsf{HD}(x)$&$\frac{1}{2}\ (\sqrt{x}-1)^2$& $\displaystyle{\sum_{a\in
      A}}\frac{1}{2}\Big(\sqrt{\mu_1(a)}-\sqrt{\mu_2(a)}\Big)^2$\\
    $\mathsf{KL}(x)$&$x\ln(x)-{x}+1$& $\displaystyle{\sum_{a\in
      A}}\mu_1(a)\ln\Big(\frac{\mu_1(a)}{\mu_2(a)}\Big)$\\
$\epsilon$-$\mathsf{D}(x)$&$\max(x-e^\epsilon,0)$ &$\displaystyle{\sum_{a\in
      A}}\max\Big({\mu_1(a)}- e^\epsilon {\mu_2(a)},0\Big)$
  \end{tabular}
 \caption{$f$-divergences for statistical distance ($\mathsf{SD}$),
   Hellinger distance ($\mathsf{HD}$), KL divergence ($\mathsf{KL}$),
   and $\epsilon$-distance ($\epsilon$-$\mathsf{D}$)}
\label{f_div}
\end{table}
Notice that some of the $\fdiv$-divergences in the table above are not
symmetric. In particular, this is the case for KL-divergence and
$\epsilon$-distance, which we use to describe $(\epsilon,\delta)$-differential
privacy. We will denote by $\mathcal{F}$ the class of functions meeting the 
requirements of \Cref{def:f-div}.

Not only do $\fdiv$-divergences measure useful statistical quantities, they also
enjoy several properties that are useful for formal verification.
(e.g.~see~\citep{csiszarS04}). 
A property that is worth mentioning and that will be used
implicitly in our example is the following.
\begin{theorem}[Data processing inequality]
\label{lem:data-processing}
  Let $\fdiv\in\mathcal{F}$, $\mu_1,\mu_2$ be two distributions over
  $A$, and $M$ be a function (potentially randomized)  mapping values in $A$ to
  distributions over $B$. Then, we have:
$$
\Delta_{\fdiv}(\mathsf{bind}\, \mu_1\, M,\mathsf{bind}\, \mu_2\,
M)\leq \Delta_{\fdiv}(\mu_1,\mu_2)
$$
\end{theorem}
Another important property for our framework is
composition. As shown by \citet{BartheO13} we can compose
\fdiv-divergences in an additive way. 
More specifically, they give the following definition.
\begin{definition}[\citet{BartheO13}]
  Let $\fdiv_1,\fdiv_2,\fdiv_3\in\mathcal{F}$. We say that
  $(\fdiv_1,\fdiv_2)$ are $\fdiv_3$ composable if and only if for
  every $A,B$, two distributions $\mu_1,\mu_2$ over $A$, and two functions
  $M_1,M_2$ mapping values in $A$ to distributions over $B$ we have
\begin{multline*}
\Delta_{\fdiv_3}(\mathsf{bind}\, \mu_1\, M_1,\mathsf{bind}\, \mu_2\,
M_2)\leq\\ \Delta_{\fdiv_1}(\mu_1,\mu_2)+\sup_v\Delta_{\fdiv_2}(M_1\,
v,M_2\, v)
\end{multline*}
\end{definition}
In particular, we have the following.
\begin{lemma}[\citet{BartheO13}]$ $
  \begin{itemize}
  \item $(\epsilon_1\text{-}\mathsf{D}, \epsilon_2\text{-}\mathsf{D})$ are
    $(\epsilon_1+\epsilon_2)\text{-}\mathsf{DP}$ composable.
  \item $(\mathsf{SD}, \mathsf{SD})$ are
    $\mathsf{SD}$ composable.
  \item $(\mathsf{HD}, \mathsf{HD})$ are
    $\mathsf{HD}$ composable.
  \item $(\mathsf{KL}, \mathsf{KL})$ are
    $\mathsf{KL}$ composable.
  \end{itemize}
\end{lemma}
This form of composition will be
internalized by the relational refinement type system that we will present
in~\Cref{sec:relational-typing}.

\section{{\large \THESYSTEM}}
\label{sec:syntax}
The main components of \THESYSTEM are a language that permits to
express Bayesian inference models and a type system for reasoning in a
relational way about programs from the language.
\subsection{The language}
The language underlying \THESYSTEM is a probabilistic programming
extension of \pcf that we will call \pcfp.
Expressions of \pcfp are defined by the following grammar
\begin{center}
  $\begin{array}{rcl}
    e
      & ::=   & x \vbar c \vbar e\ e \vbar \slam{x}{e} \\ 
      & \vbar & \sletrec{}{f}{x}{e} \vbar
                \scase{e}{\vvi{i}{\sbranch{d_i\ \vv{x_i}}{e_i}}}\\
      & \vbar &  \sunitM{e}    \vbar \sbindM{x}{e}{e}\\
&\vbar & \observe{x}{e}{e} \vbar \binfer{e} \vbar \bran{e}\\
&\vbar & \bernoulli{e} \vbar \normal{e}{e} \vbar  \betad{e}{e} \vbar
\uniform\\
 &\vbar &
 \laplace{e}{e}\vbar \gauss{e}{e}
\vbar \exponential{e}{e}{e}
  \end{array}$
\end{center}
where $c$ represents a constant from a set $\ctts$ and $x$ a variable. We will denote by \pcfp($\mathcal{X}$)
the set of expression of \THESYSTEM where the variables are taken from the set $\mathcal{X}$.
\begin{figure*}
 \begin{mathpar}
    \inferrule*[left=BindM]
     {\Gamma \vdash e_1 : \stmod{T_1}\\ 
      \Gamma, x : T_1 \vdash e_2 : \stmod{T_2}}
     {\Gamma \vdash \sbindM{x}{e_1}{e_2} : \stmod{T_2}}

     \inferrule*[left=UnitM]
     {\Gamma \vdash e : T }
     {\Gamma \vdash \sunitM{e} : \stmod{T}}
     
   \inferrule*[left=Observe]
     {\Gamma \vdash e_1 : \stmod{\coty{\tau}}  \\
      \Gamma, x : \coty{\tau} \vdash e_2 : \stmod{\stbool}}
     {\Gamma \vdash \mathsf{observe}\, x\Rightarrow e_2\, \mathsf{in}\,
  e_1 :  \stmod{\coty{\tau}}}

\inferrule*[left=Ran]
     {\Gamma \vdash e : \stdist{\coty{\tau}}}
     {\Gamma \vdash \bran{e} : \stmod{\coty{\tau}}}

   \inferrule*[left=Infer]
     {\Gamma\vdash e : \stmod{\coty{\tau}}}
     {\Gamma \vdash \binfer{e} : \stdist{\coty{\tau}}}

 \end{mathpar}
\caption{\label{fig:pcfp-typing}\pcfp type system (selected rules)}
\end{figure*}
\begin{figure}[t]
  \begin{tabular}{l}
$\mathsf{uniform}: \stdist{[0,1]}$\\
$\mathsf{bernoulli}: [0,1] \to \stdist{\stbool}$\\
$\mathsf{beta}: \rplus\times \rplus \to
    \stdist{[0,1]}$\\
$\mathsf{normal}: \reals\times \rplus \to \stdist{\R}$\\[1mm]
$\mathsf{lapMech}: \rplus\times \R \to
\stmod{\reals}$\\
$\mathsf{gaussMech}: \rplus\times \R \to \stmod{\R}$\\
$\mathsf{expMech}: \R\times ((D,R)\to \R) \times D \to
\stmod{R}$\\
        \end{tabular}
  \caption{\label{fig:pcfpdist}Primitive distributions types.}
\end{figure}

We will consider only expressions that are well typed using simple types
of the form
\begin{center}
  $\begin{array}{rcl}
    \tau,\sigma
      & ::= & \coty{\tau} \vbar
              \stmod{\coty{\tau}} \vbar               
              \stmod{\stdist{\coty{\tau}}} \vbar               
              \stdist{\coty{\tau}} \vbar
              \stfun{\tau}{\sigma}\\
    \coty{\tau}
      & ::= & \stunit \vbar \stbool \vbar \stnat \vbar \R \vbar \rplus\vbar
              \rplusinfty \vbar [0,1]\vbar
              \stlist{\coty{\tau}} .
  \end{array}$
\end{center}
where $\coty{\tau}$ are basic types. As usual a typing judgment is a
judgment of the shape $\Gamma\vdash e:\tau$ where an environment $\Gamma$ is an
assignment of types to variables. The simply typed system of \THESYSTEM is an extension of the one
in~\citet{BartheGAHRS15}; in \Cref{fig:pcfp-typing} we only present the rules specific to \THESYSTEM.

The syntax and types of \pcfp extend the one of \pcf by means of
several constructors. Basic types include the unit type $\stunit$ and
types for booleans $\stbool$ and natural numbers $\stnat$. We also
have types for real numbers $\R$, positive real numbers $\rplus$,
positive real number plus infinity $\rplusinfty$ and for real numbers
in the unit interval $[0,1]$. Finally we have lists over basic types. Simple types combines basic types using
arrow types, a probability monad $\stmod{\coty{\tau}}$ over the basic
type $\coty{\tau}$, and a type $\stdist{\coty{\tau}}$ representing
\emph{symbolic distributions} over the basic type $\coty{\tau}$.  The
probability monad can also be over symbolic distributions. 
Probabilities (\emph{actual distributions})
are encapsulated in the probabilistic monad $\stmod{\coty{\tau}}$ that can be manipulated by
the let-binder $\sbindM{x}{e_1}{e_2}$ and by the unit
$\sunitM{e}$. Symbolic distributions are built using basic probabilistic
primitives like $\bernoulli{e}$ for Bernoulli distributions,
$\normal{e_1}{e_2}$ for normal distribution, etc. These primitives are
assigned types as described in \Cref{fig:pcfpdist}.  For 
symbolic distributions we also assume that we have an operation ${\sf getParams}$ to extract the
parameters. We also have primitives
$\laplace{e_1}{e_2}$, $\gauss{e_1}{e_2}$ and
$\exponential{e_1}{e_2}{e_3}$ that provide implementations for the
mechanism ensuring differential privacy as described in
\cref{sec:dp}.  

Finally, we have three special
constructs for representing learning. The primitive
$\observe{x}{e_1}{e_2}$ can be used to describe conditional
distributions. This is a functional version of a similar primitive
used in languages like Fun~\citep{GordonABCGNRR13}. This primitive
takes two arguments, a prior $e_2$ and a predicate $e_1$ over $x$. The
intended semantics is the one provided by Bayes' theorem: it filters
the prior by means of the observation provided by $e_1$ and renormalize the
obtained distribution (see
Section~\Cref{sec:denotSemantics} for more details). The primitives $\binfer{e}$ and
$\bran{e}$ are used to transform symbolic distributions in actual
distributions and vice versa. 
In particular, $\binfer{e}$ is the main
component performing probabilistic inference.


\subsection{Denotational Semantics}
\label{sec:denotSemantics}
The semantics of \pcfp is largely standard. 
Basic types are
interpreted in the corresponding sets,
e.g. $\tyinterp{\stunit}=\{\stunit\}$, $\tyinterp{\stbool}=\{\mathsf{true},\mathsf{false}\}$,
$\tyinterp{\stnat}=\{0,1,2,\ldots\}$, etc.
As usual, arrow types $\tyinterp{\tau\to \sigma}$ are
interpreted as set of functions
$\tyinterp{\tau}\to\tyinterp{\sigma}$. A  monadic type $\stmod{\tau}$
for $\tau\in \{\coty{\tau},\stdist{\coty{\tau}}\}$ is interpreted
as the set of discrete probabilities over $\tau$, i.e.:
$$
\tyinterp{\stmod{\tau}}=
\big\{\mu:\tyinterp{\tau}\to\rplus \ |\ {\tt supp}(\mu)\, {\tt discrete}\land
\sum_{x\in\tyinterp{\tau}}\mu\, x=1 \big \}
$$

Types of the shape $\stdist{\coty{\tau}}$ are interpreted in set of
symbolic representations for distributions parametrized by values. As an example, $\stdist{\stbool}$ is interpreted as:
$$
\tyinterp{\stdist{\stbool}}=
\big\{\bernoulli{v} \ |\ v\in [0,1]\big \} 
$$
The interpretation of expressions is given as usual under a validation
$\theta$ which is a finite map from variables to values in the
interpretation of types. We will say that $\theta$ validates an
environment $\Gamma$ if $\forall x:\tau\in \Gamma$ we have
$\theta(x)\in \tyinterp{\tau}$. 
For most of the expressions the interpretation is standard, we detail the
less standard interpretations in \Cref{fig:interp}. Probabilistic
expressions are interpreted into discrete probabilities. In particular,
$\interp{\theta}{\sunitM{e}}$ is defined as the Dirac distribution
returning $\interp{\theta}{e}$ with probability one. The
binding construct $\sbindM{x}{e_1}{e_2}$ composes probabilities.
The expression  $\mathsf{observe}\, x\Rightarrow t\,  \mathsf{in}\, u$
filters the distribution $\interp{\theta}{u}$
 using the predicate $x\Rightarrow t$ and rescales it in order to obtain a
 distribution. The {\sf observe} is the key component to have conditional
 distributions and to update a prior using Bayes' theorem. The
 semantics of $\mathsf{infer}$ relies on a given algorithm\footnote{In this work we
  consider only exact inference, and we leave for future works to
  consider approximate inference. We also consider only programs with a well defined semantics: e.g. we don't consider
  programs where we make observations of events with 0 probability. We will finally restrict ourselves to inference algorithms that never fail:
 we could easily simulate this by using the $maybe$ monad.} $\mathsf{AlgInf}$ for
 inference. We leave the algorithm unspecified because it is not
 central to our verification task. 
Symbolic distributions are syntactic constructions, this is
reflected in their interpretation. For an example, we give in
\Cref{fig:interp} the interpretation
$\interp{\theta}{\bernoulli{e}}$. The operator $\mathsf{ran}$ turns a
symbolic distribution in an actual distribution. Its semantics is
defined by cases on the given symbolic distribution. In
\Cref{fig:interp} we give its interpretation for the case when the
given symbolic distribution is $\bernoulli{e}$. The cases for the
other symbolic distributions are similar. 

The soundness of the semantics is given by the following:
\begin{lemma}
  If $\Gamma\vdash e:\tau$ and $\theta$ validates $\Gamma$, then $\interp{\theta}{e}\in\tyinterp{\tau}$.
\end{lemma}
\begin{figure*}[t]
\begin{mathpar}

\interp{\theta}{\Gamma \vdash \sunitM{e} : \stmod{\tau}} =  \charfun_{\interp{\theta}{e}}
      
\interp{\theta}{\Gamma \vdash \sbindM{x}{e_1}{e_2} : \stmod{\sigma}}
 =  d \mapsto
 \displaystyle\sum_{g\in\interp{\theta}{\tau}}(\interp{\theta}{e_1}(g)\cdot
 \interp{\theta^g_x}{e_2}(d))


\interp{\theta}{\Gamma \vdash \mathsf{observe}\, x\Rightarrow t\,
  \mathsf{in}\, u: \stmod{\tau}} =  d\mapsto \frac{\interp{\theta}{u}(d) \cdot (\interp{\theta_x^{d}}{t}(\mathsf{true}))}
{\displaystyle\sum_{g\in\interp{\theta}{\tau}} ( \interp{\theta}{u}(g)
  \cdot (\interp{\theta_x^{g}}{t}(\mathsf{true})))}

\interp{\theta}{\Gamma \vdash \mathsf{infer}\, e: \stdist{\tau}}
=\mathsf{AlgInf}\,e


\interp{\theta}{\Gamma \vdash \bernoulli{e}:
  \stdist{\mathbb{B}}} =  \bernoulli{\interp{\theta}{e}}

\interp{\theta}{\Gamma \vdash \mathsf{ran}(\mathsf{bernoulli}(e)):
  \stmod{\mathbb{B}}} =   d\mapsto \begin{cases} 
          \interp{\theta}{e} & \text{if}\,d = \mathsf{true} \\
          1-\interp{\theta}{e} & \text{otherwise}
                \end{cases}



 \end{mathpar}
 \caption{\label{fig:interp}Interpretation for some of  \pcfp
   expressions (selected rules).}
\end{figure*}

\section{Relational Type System}
\label{sec:relational-typing}
\subsection{Relational Typing}
To reason about differential privacy as well as about \fdiv-divergences
we will consider a higher-order relational refinement type system. We
follow the approach proposed by~\citet{BartheGAHRS15}.

We will distinguish two sets of variables: $\rvars$ (\emph{relational}) and $\pvars$ (\emph{plain}).
Associated with every relational variable $x\in \rvars$, we have a left instance
$\l{x}$ and a right instance $\r{x}$.
We write $\rmark{\rvars}$ for $\bigcup_{x \in \rvars} \{ \l{x}, \r{x} \}$
and $\rmark{\vars}$ for $\rmark{\rvars} \cup \pvars$.

The set of \THESYSTEM~\emph{expressions} $\expr$ is the set of expressions
defined over plain variables, i.e. expressions in $\pcfp(\pvars)$.
The set of \THESYSTEM~\emph{relational expressions} $\rexpr$ is the
set of expressions defined over plain and relational variables,
expressions in $\pcfp(\rmark{\vars})$, where only non-relational 
variables can be bound.

The sets of \emph{relational types} $\rtypes = \{ T, U, \ldots \}$ and
\emph{assertions} $\rassert = \{ \phi, \psi, \ldots \}$ are defined by
the following grammars:
\begin{center}
  $\begin{array}{rrl}
    T, U \in \rtypes
      & ::= & \coty{\tau} \vbar
              \rtmod{\fdiv}{\delta}{\rtref{x}{\coty{\tau}}{\phi}}
\vbar
              \rtmod{\fdiv}{\delta}{\rtref{x}{\stdist{\coty{\tau}}}{\phi}}\\
&&
              \vbar
              \stdist{\coty{\tau}} \vbar
              \rtprod{x}{T}{T} \vbar
              \rtref{x}{T}{\phi}\\[.3em]
    \phi, \psi\in \rassert
      & ::= &
              \begin{array}[t]{@{}l@{\hspace{1cm}}l@{}}
                \fquant{\quantvar}{x}{\tau}{\phi} & (x \in \pvars)\\[.3em]
                \rquant{\quantvar}{x}{T}{\phi} & (x \in \rvars)\\[.3em]
              \end{array}\\
           && \vbar
           \Delta^{\mathfrak{D}}_\fdiv(\rmark{e},\rmark{e})\leq
           \delta\vbar
\fdiv\in\mathcal{F}\\
&& \vbar             \rmark{e} = \rmark{e} \vbar
              \rmark{e} \le \rmark{e}
\vbar \rformc(\phi_1, \ldots, \phi_n)\\[.3em]
    \rformc &= & \{ \sfrac{\rtrue}{0}, \sfrac{\rfalse}{0},
                     \sfrac{\neg}{1}, \sfrac{\vee}{2}, \sfrac{\wedge}{2},
                     \sfrac{\implies}{2} \} , 

  \end{array}$
\end{center}

\noindent where $\fdiv, \delta, \rmark{e} \in \rexpr$,
and $\quantvar\in\{\forall, \exists\}$.

Relational types extend simple types by means of relational
refinements of the shape $\rtref{x}{T}{\phi}$. This is a refinement
type that uses a relational assertion $\phi$ stating some relational
property that the inhabitants of this type have to satisfy. Relational
assertions are first order formulas over some basic predicates:
$\Delta^{\mathfrak{D}}_\fdiv(\rmark{e},\rmark{e})\leq \delta$
asserting a bound on a specific \fdiv-divergence,
$\fdiv\in\mathcal{F}$ asserting that \fdiv$\ $ is a convex function
meeting the requirements of \Cref{def:f-div}, and 
$\rmark{e} = \rmark{e}$ and $\rmark{e} \le \rmark{e}$ for the equality
and inequality of relational expressions, respectively. 
Relational types also refines the probability monad. This has now the
shape $\rtmod{\fdiv}{\delta}{\rtref{x}{T}{\phi}}$, for
$T\in\{\coty{\tau},\stdist{\coty{\tau}}\}$, and  it corresponds to a
polymonad~\citep{HicksBGLS14} or parametric effect monads~\citep{Katsumata14} where $\fdiv$ and $\delta$ are parameters useful to
reason about \fdiv-divergences. Relational expressions can appear in relational refinements
and in the parameters of the probabilistic monad, so the usual
arrow type constructor is replaced by the dependent type constructor
$\rtprod{x}{T}{S}$.

Before introducing the typing rule of \THESYSTEM we need to introduce
some notation. A relational environment $\renv{G}$ is a finite
sequence of bindings $(x :: T)$ such that each variable $x$ is never
bound twice and only relational variables from $\rvars$ are bound.  We
write $x\renv{G}$ for the type of $x$ in $\renv{G}$.  We will denote
by $\rembed{\cdot}$ the type erasure from relational types to simple
types and its extension to environments. We will use instead the
notation $\rrembed{\cdot}$, to describe the following map from
relational environments to environments $x_\svar\rrembed{\renv{G}} =
x\rembed{\renv{G}}$ iff $x\in \dom(\renv{G})$, where $\svar \in \{
\lvmark, \rvmark \}$. For example, given a relational binding $(x::T)$,
we have $\rrembed{(x::T)}= \l{x}:\rembed{T},\r{x}:\rembed{T}$.

We can now present our relational type system. \THESYSTEM proves
typing judgment of the form $\renv{G} \vdash e_1 \sim e_2 :: T$.  We
will use $\Gamma \vdash e :: T$ as a shorthand for $\Gamma \vdash e
\sim e :: T$. Several of the typing rules of \THESYSTEM come from the
system proposed by \citet{BartheGAHRS15}. We report some of them
in~\Cref{fig:relty-ext}. We also extend the subtyping relation of \citet{BartheGAHRS15}
with the rule for monadic subtyping in~\Cref{fig:subty}. 

We present the rules that are specific to \THESYSTEM
in~\Cref{fig:relty-ext}. The rules \textsf{UnitM} and \textsf{BindM}
correspond to the unit and the composition of the probabilistic
polymonad. These are similar to the usual rule for the unit and
composition of monads but additionally they require the indices to
well behave. In particular $(\fdiv_1,\fdiv_2)$ are required to be
$\fdiv_3$ composable to guarantee that the composition is
respected. The rules \textsf{Infer} and \textsf{Ran} are similar to
their simply typed version but they also transfer the information on
the \fdiv-divergence from the indices to the refinements and
viceversa. The rule \textsf{Observe} requires that the sub-expressions
are well typed relationally and it further requires the validity of the
assertion:
$$\|\renv{G}\|\vdash \Delta_\fdiv(\mathsf{observe}\, \l{x}\Rightarrow \l{e}\, \mathsf{in}\,
  \l{e'},\mathsf{observe}\, \r{x}\Rightarrow \r{e}\, \mathsf{in}\,
  \r{e'})\leq\delta''$$
for the $\delta''$ that can then be used as a bound for the
\fdiv-divergence in the conclusion. This assertion may be surprising
at first, since it doesn't consider the bounds $\delta$ and $\delta'$
for the sub-expressions but instead requires to provide directly a bound
for the conclusion---it is not really compositional. The reason for this presentation is that a
generic bound in term of $\delta$ and $\delta'$ would be often too large to
say something useful about the conclusion. In fact, estimating bounds on the
\fdiv-divergence of conditional distributions is a hard task and
often it is only expressed in term of prior perturbations~\citep{Dey1994287}. So, instead we prefer to postpone
the task to verify a bound to the concrete application where a tighter
bound can be found with some calculation. We will see
some uses of this rule in the examples in \Cref{sec:examples}.
\begin{figure}
 \begin{mathpar}
   \inferrule*[left=S-M]
   {\renv{G} \vdash T \tyle U \\
    \rrembed{\renv{G}} \vdash \delta_i : \rplusinfty \\
    \rrembed{\renv{G}} \vdash   \fdiv_i\in\mathcal{F} \\
    \forall \theta .\, \theta \VDash \renv{G},x::T \implies
    \interp{\theta}{       \fdiv_1 \le \fdiv_2
                    \wedge \delta_1   \le \delta_2   < \infty}}
   {\renv{G} \vdash \rtmod{\fdiv_1}{\delta_1}{T} \tyle \rtmod{\fdiv_2}{\delta_2}{U}}
 \end{mathpar}

 \caption{\label{fig:subty} Relational Subtyping (rule for monadic subtyping)}
\end{figure}

\begin{figure*}

\begin{mathpar}
  \inferrule*[left=UnitM]
   {\rrembed{\renv{G}} \vdash \fdiv \in \mathcal{F} \\
    \rrembed{\renv{G}} \vdash \delta   : \rplusinfty \\
    \renv{G} \vdash e :: T}
   {\renv{G} \vdash \sunitM{e} :: \rtmod{\fdiv}{\delta}{T}}

   \inferrule*[left=Infer]
     {\renv{G}\vdash e : \rtmod{\fdiv}{\delta}{\rtref{x}{\coty{\tau}}{\l{x}=\r{x}}}}
     {\renv{G} \vdash \binfer{e} : \rtref{x}{\stdist{\coty{\tau}}}{\Delta^{\mathfrak{D}}_\fdiv(\l{x},\r{x})\leq \delta} }

  \inferrule*[left=BindM]
   {\renv{G} \vdash e_1 :: \rtmod{\fdiv_1}{\delta_1}{T_1} \\ 
       (\fdiv_1,\fdiv_2)\ \text{are}\ \fdiv_3\text{-composable}     \\\\
    \renv{G} \vdash \rtmod{\fdiv_2}{\delta_2}{T_2} \\
    \renv{G}, x :: T_1 \vdash e_2 :: \rtmod{\fdiv_2}{\delta_2}{T_2}}
   {\renv{G} \vdash \sbindM{x}{e_1}{e_2} :: \rtmod{\fdiv_3}{\delta_1 + \delta_2}{T_2}}
  
\inferrule*[left=Ran]
     {\renv{G} \vdash e :
       \rtref{x}{\stdist{\coty{\tau}}}{\Delta^{\mathfrak{D}}_\fdiv(\l{x},\r{x})\leq \delta} }
     {\renv{G} \vdash \bran{e} : \rtmod{\fdiv}{\delta}{\rtref{x}{\coty{\tau}}{\l{x}=\r{x}}}}

\inferrule*[left=Observe]
{\renv{G}\vdash e:: \rtmod{\fdiv}{\delta}{\rtref{y}{\coty{\tau}}{\l{y} =\r{y}}}\\
 \renv{G}, x: \coty{\tau} \vdash e' ::\rtmod{\fdiv}{\delta'}{\rtref{y}{\mathbb{B}}{\l{y} =\r{y}}}\\
\|\renv{G}\|\vdash \Delta_\fdiv(\mathsf{observe}\, \l{x}\Rightarrow \l{e}\, \mathsf{in}\,
  \l{e'},\mathsf{observe}\, \r{x}\Rightarrow \r{e}\, \mathsf{in}\,
  \r{e'})\leq\delta''} 
{\renv{G}\vdash \mathsf{observe}\, x\Rightarrow e\, \mathsf{in}\,
  e':: \rtmod{\fdiv}{\delta''}{\rtref{y}{\coty{\tau}}{\l{y} =\r{y}}}}
 \end{mathpar}

 \caption{\label{fig:relty-ext} \THESYSTEM Relational Typing Rules}
\end{figure*}

\subsection{Relational Interpretation}
We want now to give a relational interpretation of relational types so
that we can prove the soundness of the relational type system of \THESYSTEM.
Before doing this we need to introduce an important component of our interpretation, the notion of $(\fdiv,\delta)$-lifting of a relation, inspired from the relational lifting of \fdiv-divergences by~\citet{BartheO13}.

\begin{definition}[$(\fdiv,\delta)$-Lifting of a relation $\Psi$]
\label{def:lifting} Let $\Psi\subseteq T_1\times T_2$, let
  \fdiv$\ $ be a convex function providing an \fdiv-divergence and let 
  $\delta\in\rplus$. Then, we have that $\mu_1\in\tmod{T_1}$ and
  $\mu_2\in\tmod{T_2}$ are in the $(\fdiv, \delta)$-lifting of $\Psi$,
  denoted $\lift{(\fdiv, \delta)}{\Psi}$  iff there exist
  two distributions $\mu_L,\mu_R\in\tmod{T_1\times T_2}$ such that
  \begin{enumerate}
    \item $\mu_i\, (a,b) > 0$ implies $(a,b)\in \Psi$, for $i\in\{L,R\}$,
    \item $\pi_1\, \mu_L =\mu_1\land \pi_2\, \mu_R=\mu_2$, and
    \item $\Delta_{\fdiv}(\mu_L, \mu_R)\leq \delta$.
  \end{enumerate}
where $\pi_1\, \mu = \lambda x. \sum_{y} \mu\, (x,y)$ and $\pi_2\, \mu
= \lambda y. \sum_{x} \mu\, (x,y)$.

We will call the distributions $\mu_L$ and $\mu_R$ the left and right
witnesses for the lifting, respectively.
\end{definition}
This notion of lifting will be used to give a relational interpretation of  monadic types. We say that a valuation $\theta$ validates a relational environment  $\renv{G}$, denoted $\theta \VDash \renv{G}$, if $\theta \vDash
 \rrembed{\renv{G}}$ and $\forall x \in \dom(\renv{G})$,
 $(\l{x}\theta, \r{x}\theta) \in \rinterp{\theta}{x\renv{G}}$.
The relational interpretation $\interp{\theta}{\phi} \in \{ \top, \bot \}$ of 
assertions $\phi$ with respect to a valuation $\theta \vDash \Gamma$ is an extension of the
the one provided in ~\citet{BartheGAHRS15}. In ~\Cref{fig:finterp} we provide the extensions specific to \THESYSTEM.
Notice that we interpret the assertion $\Delta_\fdiv^\mathfrak{D}(\rmark{e}_1, \rmark{e}_2)\leq \delta$ with the corresponding $\fdiv$-divergence.

We give in ~\Cref{fig:rinterp} the relational interpretation
$\rinterp{\theta}{T}$ of a relational type $T$ with respect to the
valuation $\theta \vDash \rrembed{\renv{G}}$. This corresponds to
pairs of values in the standard interpretation of \pcfp
expressions. To define this interpretation we use both the
interpretation of relational assertions given in \Cref{fig:finterp}
and the definition of lifting given in \Cref{def:lifting}. The
interpretation of relational assertions is used in the interpretation
of relational refinement types, while the lifting is used to provide
interpretation to the probabilistic polymonad. Notice that the
relational interpretation of a type $\stdist{\coty{\tau}}$ is just the
set of pairs of values in the standard interpretation of $\stdist{\coty{\tau}}$. This can then be restricted by using relational refinement types.
\begin{figure}
 \begin{mathpar}
   \inferrule
     { d_1, d_2 \in \tyinterp{\coty{\tau}} }
     { (d_1, d_2) \in \rinterp{\theta}{\coty{\tau}} }

   \inferrule
     { (d_1, d_2) \in \rinterp{\theta}{T} \\
       \interp{\rmapx{\theta}{x}{d_1}{d_2}}{\phi} }
     { (d_1, d_2) \in \rinterp{\theta}{\rtref{x}{T}{\phi}} }

   \inferrule
     { f_1, f_2 \in \tyinterp{\rembed{T} \ra \rembed{U}} \\
       \forall (d_1, d_2) \in \rinterp{\theta}{T} .\,
         (f_1(d_1), f_2(d_2)) \in \rinterp{\rmapx{\theta}{x}{d_1}{d_2}}{U} }
     { (f_1, f_2) \in \rinterp{\theta}{\rtprod{x}{T}{U}} }

   \inferrule
     { d_1, d_2\in \interp{}{\stdist{\coty{\tau}}} }
     { (d_1, d_2) \in \rinterp{\theta}{\stdist{\coty{\tau}}} }

   \inferrule
     { \mu_1, \mu_2 \in \interp{}{\stmod{\rembed{T}}} \\\\
       \lift{\fdiv, \delta}{\rinterp{\theta}{T}}\ \mu_1\ \mu_2 }
     { (\mu_1, \mu_2) \in \rinterp{\theta}{\rtmod{\fdiv}{\delta}{T}} }
 \end{mathpar}

 \caption{Relational interpretation of types \label{fig:rinterp}}
\end{figure}
We can then prove that the relational refinement type system is sound with respect to the relational interpretation of types.
\begin{theorem}[Soundness]\label{thm:soundness}
If $\renv{G} \vdash e_1 \sim e_2 :: T$, then for every valuation
$\theta\models\renv{G}$ we have
$(\interp{\theta}{e_1},\interp{\theta}{e_2})\in \rinterp{\theta}{T}$.
\end{theorem}
The soundness theorem above give us a concrete way to reason about \fdiv-divergences. 
\begin{corollary}[\fdiv-divergence]
  If $\vdash e ::  \rtmod{\fdiv}{\delta}{\rtref{y}{\tau}{\l{y}=\r{y}}}$
  then for every $(\mu_1,\mu_2)\in\interp{}{e}$ we have $\Delta_{\fdiv}(\mu_1,\mu_2)\leq \delta$.
\end{corollary}

Moreover, thanks to the characterization of differential privacy in terms of \fdiv-divergence given by~\citet{BartheO13} we can refine the previous result to show that \THESYSTEM accurately models differential privacy.
\begin{corollary}[Differential Privacy]
  If $\vdash e :: \rtref{x}{\sigma}{\Phi} \rightarrow
  \mathfrak{M}_{\epsilon\textnormal{-}\mathsf{D},\delta}{\rtref{y}{\tau}{\l{y}=\r{y}}}$
  then $\interp{}{e}$ is $(\epsilon,\delta)$-differentially private
  w.r.t.\, adjacency relation $\interp{}{\Phi}$.
\end{corollary}
\begin{figure}
  \begin{center}
    $\begin{array}{r@{\;}c@{\;}l}
    
      \interp{\theta}{\fdiv\in\mathcal{F}} & = &
        \interp{\theta}{\fdiv}\in\mathcal{F}\\[.5em]
        \interp{\theta}{\Delta_\fdiv^\mathfrak{D}(\rmark{e}_1, \rmark{e}_2)\leq \delta} & = &
        \Delta_{\interp{\theta}{\fdiv}}(\interp{\theta}{\rmark{e}_1},
        \interp{\theta}{\rmark{e}_2})\leq \interp{\theta}{\delta}\\[.5em]
    
    \end{array}$
  \end{center}

 \caption{\label{fig:finterp} Relational interpretation of assertions (added rules)}
\end{figure}

\section{Examples}
\label{sec:examples}
In this section we show how we can use \THESYSTEM to guarantee
differential privacy for Bayesian learning by adding noise on the
input, noise on the output using $\ell_1$ norm, and noise on the
output using \fdiv-divergences.  
We will show some of these approaches on three classical examples
from Bayesian learning: 
learning the bias of a coin from some observations (as discussed in
\Cref{sec:motivations}), its generalized process, i.e the Dirichlet/multinomial model and the learning of the mean of a Gaussian. In all the example we will use
pseudo code that can be easily desugared into the language presented in
\Cref{sec:syntax}. Indeed, the following examples have been type-checked
with an actual tool implementing \THESYSTEM. More examples can be found in the supplementary material section.
\subsection{Input perturbation}
\paragraph{Input perturbation: Beta Learning}
Let's start by revisiting the task of inferring the parameter of a
Bernoulli distributed random variable given a sequence of private
observations. 
We consider two lists of booleans with the same length 
in the adjacency relation $\Phi$ iff they differ in the value of at
most one entry. 
We want to ensure differential privacy by perturbing the input. A
natural way to do this, since the observations are boolean value is by
using the exponential mechanism. We can  then learn the bias from the perturbed data. The post-processing property of differential privacy
ensures that we will learn the parameter in a private way.

Let's start by considering the quality score function for the exponential mechanism. A natural choice is to consider a function  \texttt{score:bool}$\rightarrow$\texttt{bool}$\rightarrow$\texttt{\{0,1\}} mapping equal booleans to 1 and different booleans to 0. Remember that the intended reading is that one of the boolean is the one to which we want to give a quality score, while the other is the one provided by the observation.  The sensitivity of \texttt{score} is 1.
Using this score function we can then create a general function for adding noise to the input list:
\begin{Verbatim}[commandchars=\\\{\}]
1.  \sf {let rec} addNoise db eps  = {match} db {with}
2.  \sf | []    \myarr {return} ([])
3.  \sf | y::yl \myarr {mlet} yn = ({expMech} eps score y) {in}
4.  \sf             {mlet} yln = (addNoise yl eps) {in} {return}(yn::yln)
\end{Verbatim}
To this function we can give the following type guaranteeing differential privacy.
$$
\{l:: \stlist{\stbool} \mid \l{l}\ \Phi\ \r{l}\}\rightarrow
\{\epsilon::\rplus \mid =\}\rightarrow
\mathfrak{M}_{\epsilon\textnormal{-}\mathsf{D},0}\{b::\stlist{\stbool} \mid = \}
$$
where we use $\{x::T \mid = \}$ as a shorthand for  $\{x::T \mid  \l{x}=\r{x} \}$. We will use this shorthand all along this section.

The bulk of the example is the following function that recursively
updates the prior distribution and learns the final distribution over
the parameter. 
\begin{Verbatim}[commandchars=\\\{\}]
1. \sf let rec learnBias dbn prior = match dbn with
2. \sf | []     \myarr prior
3. \sf | d::dbs \myarr \textsf{observe} 
4. \sf        (\textsf{fun} r \myarr \textsf{mlet} z = {ran} {bernoulli}(r) \textsf{in} \textsf{return} (d=z))
5. \sf        (learnBias dbs prior)
\end{Verbatim}
The likelihood given in Line 4 is the formal version of the one we presented in \Cref{sec:motivations}.
The function {\sf learnBias} can be typed in different ways depending on what is our goal. For this example we can assign to it the following type:
$$
\{l::\stlist{\stbool} \mid = \}\rightarrow
\mathfrak{M}_{\mathsf{SD},0}\{x:: [0,1] \mid =\}\rightarrow
\mathfrak{M}_{\mathsf{SD},0}\{x:: [0,1] \mid = \}
$$

The reading of this type is that if ${\sf learnBias}$ takes two lists
of observations that are equal and two prior that are equal, then we
obtain two posterior that are equal. Thanks to this we can type the occurrence
of {\sf observe} in line 3-4 using a trivial assertion. Here we use the {\sf SD} divergence but in fact this would also hold for any other $\fdiv\in\mathcal{F}$. In particular, this type allows us to compose it with {\sf addNoise} using an {\sf mlet}. This type also reflects the fact that the prior is public.
We can then compose these two procedures in the following program:
\begin{Verbatim}[commandchars=\\\{\}]
1. \sf {let} main db a b eps = {mlet} noisyDB = (addNoise db eps) 
2. \sf   {in} {return}({infer} (learnBias noisyDB ({ran} ({beta}(a,b)))))
\end{Verbatim}
Notice that in line 2 we use {\sf infer} for learning from the noised data. 
We can then assign to {\sf main} the type 
\begin{multline*}
\{l::\ \stlist{\stbool} \mid \l{l}\ \Phi\ \r{l}\}\rightarrow\{a::\rplus \mid =
\}\rightarrow \{b::\rplus\ \mid = \}\rightarrow\\
\{\epsilon:: \rplus \mid = \}\rightarrow
\mathfrak{M}_{\epsilon\textnormal{-}\mathsf{D},0}\{d:: \mathfrak{D} [[0,1]] \mid = \}
\end{multline*}
which guarantees us that the result is $\epsilon$ differentially private. Notice that the result type is a polymonadic type over $\mathfrak{D} [[0,1]]$. This because we are releasing the symbolic distribution.
\paragraph{Input perturbation: Normal Learning}
An example similar to the previous one is learning the mean of a
gaussian distribution with known variance: \verb kv,   from a list of
real number observations---for instance some medical parameters like the level
of LDL of each patient. We consider two lists of reals with the same
length adjacent when the $\ell_1$ distance between at most two elements (in the same
position) is bounded by 1. To perturb the input we may now want to use
a different mechanism, for example we could use the Gaussian
mechanisms---this may give reasonable results if we expect the data to come from a normal
distribution.  Also in this case, the sensitivity is 1. 
The \textsf{addNoise}  function is very similar to the one we used in
the previous example:
\begin{Verbatim}[commandchars=\\\{\}]
1.  \sf {let rec} addNoise db eps delta = {match} db {with}
2.  \sf | []    \myarr {return} ([])
3.  \sf | y::yl \myarr {mlet} yn = ({gaussMech} (sigma eps delta) y) {in}
4.  \sf        {mlet} yln = (addNoise yl eps delta) {in} {return}(yn::yln)
\end{Verbatim}
The two differences are that now we also have \textsf{delta} as input
and that in line 3 instead of the score function we have a function
\textsf{sigma} computing the variance as in \Cref{def:gaussMech}.
The inference function become instead the following.
\begin{Verbatim}[commandchars=\\\{\}]
1. \sf \textsf{let rec} learnMean dbn prior = \textsf{match} dbn \textsf{with}
2.   \sf| []     \myarr prior
3.   \sf| d::dbs \myarr \textsf{observe} (\textsf{fun} (r: real) \myarr
4.     \sf             \textsf{mlet} z = \textsf{ran} \textsf{normal}(r, kv) \textsf{in} \textsf{return} (d=z))
5.            \sf           (learnMean dbs prior) \textsf{in}
6. \sf \textsf{let} main db hMean hVar eps delta =
7.     \sf \textsf{mlet} noisyDB = (addNoise db eps delta)\textsf{in}
8.        \sf \textsf{return}(\textsf{infer} (learnMean noisyDB 
9.           \sf          (\textsf{ran} (\textsf{normal}(hMean,hVar)))))
\end{Verbatim}

Composing them we get the following type guaranteeing $(\epsilon,\delta)$-differential privacy.
\begin{multline*}
\{l::\stlist{\streal}\mid \l{l}\ \Phi\ \r{l}\}\rightarrow\{a::\rplus\mid  = \}\rightarrow \{b::\rplus\mid = \}\rightarrow\\
\{\epsilon:: \rplus\mid = \}\rightarrow \{\delta:: \rplus\mid = \}\rightarrow\mathfrak{M}_{\epsilon\textnormal{-}\mathsf{D},\delta}\{d:: \mathfrak{D}[\streal]\mid = \}
\end{multline*}
\subsection{Noise on Output with $\ell_1$-norm}
We present  examples where the privacy guarantee is achieved by
adding noise on the output. For doing this we need to compute the sensitivity of the program. 
In contrast, in the previous section the sensitivity was evident
because directly computed on the input. As discussed before we can
compute the sensitivity with respect to different metrics. Here we consider the sensitivity computed over the $\ell_{1}$-norm on the parameters 
of the posterior distribution.
\paragraph{Output parameters perturbation: Beta Learning}
The main difference with the example in the previous section is that here we add Laplacian noise to 
the parameters of the posterior. 
\begin{Verbatim}[commandchars=\\\{\}]
1. \sf\textsf{let} main db a b eps=
2.  \sf\textsf{let}  d = \textsf{infer} (learnBias db (ran beta(a,b))) \textsf{in}
3.   \sf\textsf{let} (aP, bP) = \textsf{getParams} d \textsf{in}
4.   \sf\textsf{mlet} aPn = \textsf{lapMech}(eps, aP) in
5.   \sf\textsf{mlet} bPn = \textsf{lapMech}(eps, bP) in
6.   \sf\textsf{return} \textsf{beta}(aPn, bPn)
\end{Verbatim}
In line 2 we use the function \textsf{learnBias} from the previous
section, while in line 4 and 5 we add Laplace noise.  The formal sensitivity
analysis is based on the fact that the posterior parameters are going
to be the counts of \emph{true} and \emph{false} in the data
respectively summed up to the parameters of the prior. This reasoning
is performed on each step of {\sf observe}. Then we can prove that
the $\ell_1$-norm sensitivity of the whole program is 2 and  type the
program with a type guaranteeing $2\epsilon$-differentially privacy.
\begin{flalign*} 
\{l:: \stlist{\stbool} \mid\l{l}\ \Phi\ \r{l}\}\rightarrow\{a:: \rplus \mid =
\}\rightarrow\{b:: \rplus  \mid  = \}\rightarrow\\
\{\epsilon:: \rplus  \mid =
\}\rightarrow\mathfrak{M}_{2\epsilon\textnormal{-}\mathsf{D},0}\{d::
\mathfrak{D}[[0,1]] \mid  = \}
\end{flalign*}
\paragraph{Output parameters perturbation: Normal Learning}
For this example we use the same adjacency relation of the example
with noise on the input where in particular the number of observation
$n$ is public knowledge. The code is very similar to the previous one. 
\begin{Verbatim}[commandchars=\\\{\}]
1.\sf\textsf{let} main db hM hV kV eps =
2. \sf \textsf{let} mDistr = \textsf{infer} \sf(learnMean db (\textsf{ran} \textsf{normal}(hM,kV))) \textsf{in}
3.  \sf\textsf{let}  mean = \textsf{getMean} mDistr \textsf{in}
4.  \sf\textsf{mlet} meanN = \textsf{lapMech}(eps/s mean) in
5.  \sf\textsf{let} d = \textsf{normal}(meanN, uk) \textsf{in} return(d)
\end{Verbatim}
where $uk=\big(\frac{1}{hV^2}+\frac{n}{kv^2}\big)^{-1}$. Notice that we only add noise to the posterior 
mean parameter and not to the posterior variance parameter since the latter doesn't depend on the data but only on public information.
The difficulty for verifying this example is in the sensitivity analysis. By
some calculations this can be bound by
$s=\frac{hV}{ kv + hV}$ where $kv$ is the known variance of the gaussian distribution whose mean we are learning and 
$hV$ is the prior variance over the mean. We use this information in line 4 when we add
noise with the Laplace mechanism. By using
this information we can give the following type to the previous program:
\begin{multline*} 
\{l:: \stlist{\reals} \mid\l{l}\ \Phi\ \r{l} \} \rightarrow\{hM:: \reals \mid  =
\}\rightarrow\{hV:: \rplus  \mid = \}\rightarrow\\\{kv:: \rplus  \mid =
\}\rightarrow\{\epsilon:: \rplus  \mid =
\}\rightarrow\mathfrak{M}_{s\epsilon\textnormal{-}\mathsf{D},0}\{d::
\mathfrak{D}[\real] \mid  = \}
\end{multline*}
\subsection{Noise on Output using \fdiv-divergences}
We now turn to the approach of calibrating the sensitivity according
to \fdiv-divergences.  We will consider once again the example for
learning privately the distribution over the parameter of a Bernoulli
distribution, but differently from the previous section we will add
noise to the output of the inference algorithm using the exponential
mechanism with a score function using an \fdiv-divergence. So, we 
perturb the output distribution and not its parameters.

We will use Hellinger distance as a metric over the output space of our
differentially private program, but any other \fdiv-divergence could
also be used.
The quality score function for the exponential mechanism can be given a
type of the shape:
$$
\{l:: \stlist{\stbool}\mid \l{l}\ \Phi\ \r{l} \}
\rightarrow\{d:: \mathfrak{D}[\tau] \mid = \}
\rightarrow\{r::\reals \mid |\l{r}-\r{r}|\leq \rho\}
$$
where the bound $\rho$ express its sensitivity. Now we can use as a
score function the following program
\begin{Verbatim}[commandchars=\\\{\}]
\sf score (db, prior) out = -(H (\textsf{infer} (learnBias db prior)) out)
\end{Verbatim}
This quality score uses a function {\sf H} computing the Hellinger
distance between the result of the inference and a potential output to
assign it a score.
The closer {\sf out}  is to
the real distribution (using Hellinger distance), the higher
the scoring is. If we use the exponential mechanism with this
score we achieve our goal of using the Hellinger to ``calibrate
the noise''. Indeed we have a program:
\begin{Verbatim}[commandchars=\\\{\}]
\sf \textsf{let} main prior obs eps = \textsf{expMech} eps score (obs, prior)
\end{Verbatim}
To which we can assign type:
\begin{multline*} 
\mathfrak{M}_{\mathsf{HD},0}\{x:: [0,1] \mid =
 \}\rightarrow\{\ell::\stlist{\stbool} \mid \l{\ell}\ \Phi\
 \r{\ell}\}\\ \rightarrow\{\epsilon:: \rplus \mid = \}\rightarrow
\mathfrak{M}_{\rho\epsilon\textnormal{-}\mathsf{D},0}\{ d::\mathfrak{D}
 [[0,1]]  \mid = \}
\end{multline*}
Concretely, to achieve this we can proceed by considering first the code for \textsf{learnBias}:
\begin{Verbatim}[commandchars=\\\{\}]
1.\sf\textsf{let rec} learnBias db prior = \textsf{match} dbn \textsf{with}
2.\sf| []     \myarr prior
3.\sf| d::dbs \myarr mlet rec = (learBias dbs prior) in \textsf{observe}
4.  (\textsf{fun} r \myarr \sf\textsf{mlet} z = \textsf{ran} \textsf{bernoulli}(r) \textsf{in} \textsf{return} (d=z)) rec
\end{Verbatim}
To have a bound for the whole {\sf learnBias} we need first to give a
bound to the difference in Hellinger distance that two distinct observations can
generate. This is described by the following lemma.
\begin{lemma}
\label{hdlemma}
Let $d_1, d_2: \stbool$ with $d_1 \Phi d_2$. Let $a,b \in \rplus$. Let
$\Pr(\xi)={\sf Beta}(a,b)$. Then
$\Delta_{\mathsf{HD}}(\Pr(\xi \mid d_1), \Pr(\xi \mid d_2))\leq
\sqrt{1-\frac{\pi}{4}}=\rho$.
\end{lemma}
Using this lemma we can then type the {\sf observe} statement with the
bound $\rho$. 
We still need to propagate this bound to the whole {\sf learnBias}. We
can do this by using the adjacency relation which imposes at most one
difference in the observations, and the data processing inequality
\Cref{lem:data-processing} guaranteeing that for equal observations the
Hellinger distance cannot increase.
Summing up,  using the lemma above, the adjacency assumption and the data
processing inequality  we can give to {\sf learnBias} the following type:   
\begin{multline*} 
\{l:: \stlist{\stbool} \mid\l{l}\ \Phi\ \r{l}\}\rightarrow\mathfrak{M}_{\mathsf{HD},
  0}\{x:: [0,1] \mid = \}\\
\rightarrow\mathfrak{M}_{\mathsf{HD}, \rho}\{x:: [0,1]
\mid= \}
\end{multline*}
This ensures that starting from the same
prior and observing $l_1$ and $l_2$ in the two different runs such that
$l_1\ \Phi\ l_2$ 
we can achieve two beta distributions which are at distance at most $\rho$.
Using some additional refinement for {\sf infer} and {\sf H} we can guarantee that
{\sf score} has the intended type, and so we can guarantee that
overall this program is 
$(\rho\epsilon,0)$-differential privacy. 

The reasoning above is not limited to the Hellinger distance. For instance the following lemma:
\begin{lemma}
Let $d_1, d_2: \stbool$ with $d_1 \Phi d_2$. Let $a,b \in \rplus$. Let
$\Pr(\xi)={\sf Beta}(a,b)$. Then
$\Delta_{\mathsf{SD}}(\Pr(\xi \mid d_1), \Pr(\xi \mid d_2))\leq
\sqrt{2(1-\frac{\pi}{4})}=\zeta$.
\end{lemma}
\noindent gives a type in term
of statistical distance:
\begin{multline*} 
\{l:: \stlist{\stbool} \mid\l{l}\ \Phi\ \r{l}\}\rightarrow\mathfrak{M}_{\mathsf{SD},
  0}\{x:: [0,1] \mid = \}\\
\rightarrow\mathfrak{M}_{\mathsf{SD}, \zeta}\{x:: [0,1]
\mid= \}
\end{multline*}
The choice of which metric to use is ultimately left to the user.
This example easily extends also to the Dirichlet example. More details about the Dirichlet distribution are
given in the supplementary material section.
Indeed, Lemma \ref{hdlemma} can be generalized to arbitrary Dirichlet distributions:
\begin{lemma}
Let $k\in \mathbb{N}^{\geq2}$, $d_1, d_2: \stlist{ [k]}$ with $d_1 \Phi d_2$. \\Let $a_1,a_2,\dots,a_k\in \rplus$. Let
$\Pr(\xi)={\sf Dirichlet}(a_1,a_2,\dots,a_k)$. Then
$\Delta_{\mathsf{HD}}(\Pr(\xi \mid d_1), \Pr(\xi \mid d_2))\leq
\sqrt{1-\frac{\pi}{4}}=\rho$.
\end{lemma}
Using this lemma we can assign to the following program:
\begin{Verbatim}[commandchars=\\\{\}]
1.\sf\textsf{let rec} learnP db prior = \textsf{match} dbn \textsf{with}
2.\sf| []     \myarr prior
3.\sf| d::dbs \myarr mlet rec = (learnP dbs prior) in \textsf{observe}
4.  (\textsf{fun} r s\myarr \sf\textsf{mlet} z = \textsf{ran} \textsf{multinomial}(r,s) \textsf{in}
5.                           \textsf{return} (d=z)) rec
\end{Verbatim}
the type:
\begin{multline*} 
\{l:: \stlist{[\textbf{3}]} \mid\l{l}\ \Phi\ \r{l}\}\rightarrow\mathfrak{M}_{\mathsf{HD},
  0}\{x:: [0,1]^2 \mid = \}\\
\rightarrow\mathfrak{M}_{\mathsf{HD}, \rho}\{x:: [0,1]^2
\mid= \}
\end{multline*}
Similarly to the previous example we can now add noise to the output
of the inference process using the sensitivity with respect to the
Hellinger distance and obtain a $(\rho\epsilon,0)$-differential
privacy guarantee. 
\section{Related work}

\paragraph*{Differential privacy and Bayesian inference}

Our system targets programs from the combination of differential privacy and
Bayesian inference. Both of these topics are active areas of research, and their
intersection is an especially popular research direction today. We briefly
summarize the most well-known work, and refer interested readers to surveys for
a more detailed development (\citet{DBLP:journals/fttcs/DworkR14} for
differential privacy, \citet{Bishop:2006} for Bayesian inference).

\citet{DBLP:conf/pods/BlumDMN05} and \citet{DMNS06} proposed \emph{differential privacy}, a
worst-case notion of statistical privacy, in a pair of groundbreaking papers,
initiating intense research interest in developing differentially private
algorithms. The original works propose the Laplace and Gaussian mechanisms that
we use, while the seminal paper of \citet{MT07} introduces the exponential
mechanism.
Recently, researchers have investigated how to guarantee differential privacy
when performing \emph{Bayesian inference}, a foundational technique in machine
learning. 
Roughly speaking, works in the literature have explored three different approaches to guaranteeing
differential privacy when the samples are private data. First, we may
add noise directly to the samples, and then perform inference as usual
\citep{DBLP:conf/nips/WilliamsM10}. Second, we may perform inference
on the private data, then add noise to the parameters themselves
\citep{ZhangRD16}. This approach requires bounding the sensitivity of the
output parameters when we change a single data sample, relying on
specific properties of the model and the prior distribution. The final
approach involves no noise during inference, but outputs samples from
the posterior rather than the entire posterior distribution
\citep{DimitrakakisNMR14,ZhangRD16,Zheng16}. This last approach is highly specific to the model and
prior, and our system does not handle this method of achieving
privacy, yet.

\paragraph*{Formal verification for differential privacy}

In parallel with the development of private algorithms, researchers in formal
verification have proposed a wide variety of techniques for \emph{verifying}
differential privacy. For a comprehensive discussion, interested readers can
consult the recent survey by \citet{Barthe:2016:PLT}. Many of these techniques rely on the composition
properties of privacy, though there are some exceptions \citep{BGGHS16}.  For a
brief survey, the first systems were based on runtime verification of privacy
\citep{PINQ09}. The first systems for static verification of privacy used linear
type systems \citep{ReedPierce10,GaboardiHHNP13}. There is also extensive work
on relational program logics for differential privacy
\citep{POPL:BKOZ12,DBLP:conf/csfw/BartheDGKB13,BartheO13}, and techniques for
verifying privacy in standard Hoare logic using product programs
\citep{BGGHKS14}. None of these techniques have been applied to verifying differential privacy of Bayesian inference.
Our system is most closely related to $\mathsf{HOARe}^2$, a relational
refinement type system that was recently proposed by \citet{BartheGAHRS15}. This
system has been used for verifying differential privacy of algorithms, and more
general relational properties like incentive compatibility from the field of
mechanism design. However, it cannot model probabilistic inference.

\paragraph*{Probabilistic programming}
Research in probabilistic programming has emerged early in the 60s and
70s, and is nowadays a very active research area. 
%
Relevant to our work is in particular the research in probabilistic programming
for machine learning and statistics which has been very active in
recent years. Many probabilistic programming languages have been
designed for these applications, including WinBUGS~\citep{LunnTBS00},
IBAL~\citep{Pfeffer01}, Church~\citep{DBLP:conf/uai/GoodmanMRBT08},
Infer.net~\citep{InferNET12}, Tabular~\citep{GordonGRRBG14},
Anglican~\citep{TolpinMW15}, Dr. Bayes~\citep{TorontoMH15}. 
Our goal is not
to provide a new language but instead is to propose a framework where
one can reason about differential privacy for such languages.
For instance, we compiled programs written in Tabular ~\citep{GordonGRRBG14}  into \THESYSTEM
so that differential privacy could be verified. More information on this translation can be found in the supplementary material section.
Another related work is the one by \citet{AdamsJ15} proposing a type theory for Bayesian inference. While
technically their work is very different from ours it shares the 
same goal of providing reasoning principles for Bayesian inference.
Our work considers a probabilistic \pcf for discrete distributions.
It would be interesting to extend our techniques to higher-order languages with continuous distributions and conditioning,
by building on the rigorous foundations developed in recent work \cite{lics16,icfp16}.

\section{Conclusion}
We have presented \THESYSTEM, a type-based framework for differentially private
Bayesian inference. Our framework allows to write data analysis as
functional programs for Bayesian inference and add to noise to them in different
ways using different metrics. Besides, our framework allows to reason about general
\fdiv-divergences for Bayesian inference. 

Future directions include exploring the use of this approach to
guarantee robustness for Bayesian inference and other machine learning
techniques~\citep{Dey1994287}, to ensure differential privacy using conditions
over the prior and the likelihood similar to the ones studied by
\citet{Zheng16,ZhangRD16}, and investigating further uses of \fdiv-divergences for
improving the utility of differentially private Bayesian
learning. On the programming language side it would also be
interesting to extend our framework to continuous distributions
following the approach by~\citet{sato2016}. We
believe that the intersection of programming languages, machine
learning, and differential privacy will reserve us many exciting results.

\bibliographystyle{abbrvnat}
\bibliography{newbib}

\begin{thebibliography}{42}
\providecommand{\natexlab}[1]{#1}
\providecommand{\url}[1]{\texttt{#1}}
\expandafter\ifx\csname urlstyle\endcsname\relax
  \providecommand{\doi}[1]{doi: #1}\else
  \providecommand{\doi}{doi: \begingroup \urlstyle{rm}\Url}\fi

\bibitem[{Adams} and {Jacobs}(2015)]{AdamsJ15}
R.~{Adams} and B.~{Jacobs}.
\newblock A type theory for probabilistic and bayesian reasoning.
\newblock \emph{CoRR}, abs/1511.09230, 2015.

\bibitem[Barthe and Olmedo(2013)]{BartheO13}
G.~Barthe and F.~Olmedo.
\newblock Beyond differential privacy: Composition theorems and relational
  logic for f-divergences between probabilistic programs.
\newblock In \emph{ICALP}, 2013.

\bibitem[Barthe et~al.(2012)Barthe, K\"{o}pf, Olmedo, and
  {Zanella-B{\'e}guelin}]{POPL:BKOZ12}
G.~Barthe, B.~K\"{o}pf, F.~Olmedo, and S.~{Zanella-B{\'e}guelin}.
\newblock Probabilistic {R}elational {R}easoning for {D}ifferential {P}rivacy.
\newblock In \emph{POPL}, 2012.

\bibitem[Barthe et~al.(2013)Barthe, Danezis, Gr{\'e}goire, Kunz, and
  Zanella~B{\'e}guelin]{DBLP:conf/csfw/BartheDGKB13}
G.~Barthe, G.~Danezis, B.~Gr{\'e}goire, C.~Kunz, and S.~Zanella~B{\'e}guelin.
\newblock Verified computational differential privacy with applications to
  smart metering.
\newblock In \emph{CSF}, 2013.

\bibitem[Barthe et~al.(2014)Barthe, Gaboardi, Gallego~Arias, Hsu, Kunz, and
  Strub]{BGGHKS14}
G.~Barthe, M.~Gaboardi, E.~J. Gallego~Arias, J.~Hsu, C.~Kunz, and P.-Y. Strub.
\newblock Proving differential privacy in {H}oare logic.
\newblock In \emph{CSF}, 2014.

\bibitem[Barthe et~al.(2015)Barthe, Gaboardi, Arias, Hsu, Roth, and
  Strub]{BartheGAHRS15}
G.~Barthe, M.~Gaboardi, E.~J.~G. Arias, J.~Hsu, A.~Roth, and P.~Strub.
\newblock Higher-order approximate relational refinement types for mechanism
  design and differential privacy.
\newblock In \emph{POPL}, 2015.

\bibitem[Barthe et~al.(2016{\natexlab{a}})Barthe, Gaboardi, Gr{\'e}goire, Hsu,
  and Strub]{BGGHS16}
G.~Barthe, M.~Gaboardi, B.~Gr{\'e}goire, J.~Hsu, and P.-Y. Strub.
\newblock Proving differential privacy via probabilistic couplings.
\newblock In \emph{LICS}, 2016{\natexlab{a}}.

\bibitem[Barthe et~al.(2016{\natexlab{b}})Barthe, Gaboardi, Hsu, and
  Pierce]{Barthe:2016:PLT}
G.~Barthe, M.~Gaboardi, J.~Hsu, and B.~Pierce.
\newblock Programming language techniques for differential privacy.
\newblock \emph{ACM SIGLOG News}, 2016{\natexlab{b}}.

\bibitem[Bishop(2006)]{Bishop:2006}
C.~M. Bishop.
\newblock \emph{Pattern Recognition and Machine Learning (Information Science
  and Statistics)}.
\newblock 2006.
\newblock ISBN 0387310738.

\bibitem[Blum et~al.(2005)Blum, Dwork, McSherry, and
  Nissim]{DBLP:conf/pods/BlumDMN05}
A.~Blum, C.~Dwork, F.~McSherry, and K.~Nissim.
\newblock Practical privacy: {T}he {SuLQ} framework.
\newblock In \emph{PODS}, 2005.

\bibitem[Borgstr{\"o}m et~al.(2016)Borgstr{\"o}m, Lago, Gordon, and
  Szymczak]{icfp16}
J.~Borgstr{\"o}m, U.~D. Lago, A.~D. Gordon, and M.~Szymczak.
\newblock A lambda-calculus foundation for universal probabilistic programming.
\newblock In \emph{ICFP}, 2016.

\bibitem[Chaudhuri et~al.(2011)Chaudhuri, Monteleoni, and
  Sarwate]{Chaudhuri:2011}
K.~Chaudhuri, C.~Monteleoni, and A.~D. Sarwate.
\newblock Differentially private empirical risk minimization.
\newblock 2011.

\bibitem[Csisz\'ar(1963)]{csiszar63}
I.~Csisz\'ar.
\newblock Eine informationstheoretische {U}ngleichung und ihre {A}nwendung auf
  den {B}eweis der {E}rgodizitat von {M}arkoffschen {K}etten.
\newblock \emph{Magyar. Tud. Akad. Mat. Kutat\'o Int. K\"ozl}, 1963.

\bibitem[Csisz\'ar and Shields(2004)]{csiszarS04}
I.~Csisz\'ar and P.~Shields.
\newblock Information theory and statistics: A tutorial.
\newblock \emph{Foundations and Trends in Communications and Information
  Theory}, 2004.

\bibitem[Dey and Birmiwal(1994)]{Dey1994287}
D.~K. Dey and L.~R. Birmiwal.
\newblock {R}obust {B}ayesian analysis using divergence measures.
\newblock \emph{Statistics \& Probability Letters}, 1994.

\bibitem[Dimitrakakis et~al.(2014)Dimitrakakis, Nelson, Mitrokotsa, and
  Rubinstein]{DimitrakakisNMR14}
C.~Dimitrakakis, B.~Nelson, A.~Mitrokotsa, and B.~I.~P. Rubinstein.
\newblock Robust and {P}rivate {B}ayesian {I}nference.
\newblock In \emph{ALT}, 2014.

\bibitem[Dwork and Roth(2014)]{DBLP:journals/fttcs/DworkR14}
C.~Dwork and A.~Roth.
\newblock The algorithmic foundations of differential privacy.
\newblock \emph{Foundations and Trends in Theoretical Computer Science}, 2014.

\bibitem[Dwork et~al.(2006)Dwork, McSherry, Nissim, and Smith]{DMNS06}
C.~Dwork, F.~McSherry, K.~Nissim, and A.~Smith.
\newblock Calibrating noise to sensitivity in private data analysis.
\newblock In \emph{TCC}, 2006.

\bibitem[Dwork et~al.(2010)Dwork, Rothblum, and Vadhan]{DworkRV10}
C.~Dwork, G.~N. Rothblum, and S.~P. Vadhan.
\newblock Boosting and differential privacy.
\newblock In \emph{FOCS}, 2010.

\bibitem[Ebadi et~al.(2015)Ebadi, Sands, and Schneider]{conf/popl/EbadiSS15}
H.~Ebadi, D.~Sands, and G.~Schneider.
\newblock Differential privacy: Now it's getting personal.
\newblock \emph{POPL}, 2015.

\bibitem[Eigner and Maffei(2013)]{EignerM13}
F.~Eigner and M.~Maffei.
\newblock Differential privacy by typing in security protocols.
\newblock In \emph{CSF}, 2013.

\bibitem[Gaboardi et~al.(2013)Gaboardi, Haeberlen, Hsu, Narayan, and
  Pierce]{GaboardiHHNP13}
M.~Gaboardi, A.~Haeberlen, J.~Hsu, A.~Narayan, and B.~C. Pierce.
\newblock Linear dependent types for differential privacy.
\newblock In \emph{POPL}, 2013.

\bibitem[Goodman et~al.(2008)Goodman, Mansinghka, Roy, Bonawitz, and
  Tenenbaum]{DBLP:conf/uai/GoodmanMRBT08}
N.~D. Goodman, V.~K. Mansinghka, D.~M. Roy, K.~Bonawitz, and J.~B. Tenenbaum.
\newblock Church: a language for generative models.
\newblock In \emph{UAI}, 2008.

\bibitem[Gordon et~al.(2013)Gordon, Aizatulin, Borgstr{\"{o}}m, Claret,
  Graepel, Nori, Rajamani, and Russo]{GordonABCGNRR13}
A.~D. Gordon, M.~Aizatulin, J.~Borgstr{\"{o}}m, G.~Claret, T.~Graepel, A.~V.
  Nori, S.~K. Rajamani, and C.~V. Russo.
\newblock A model-learner pattern for bayesian reasoning.
\newblock In \emph{POPL}, 2013.

\bibitem[Gordon et~al.(2014)Gordon, Graepel, Rolland, Russo, Borgstr{\"{o}}m,
  and Guiver]{GordonGRRBG14}
A.~D. Gordon, T.~Graepel, N.~Rolland, C.~V. Russo, J.~Borgstr{\"{o}}m, and
  J.~Guiver.
\newblock Tabular: a schema-driven probabilistic programming language.
\newblock In \emph{POPL}, 2014.

\bibitem[Hardt et~al.(2012)Hardt, Ligett, and McSherry]{HardtLM12}
M.~Hardt, K.~Ligett, and F.~McSherry.
\newblock A simple and practical algorithm for differentially private data
  release.
\newblock In \emph{NIPS}, 2012.

\bibitem[Hicks et~al.(2014)Hicks, Bierman, Guts, Leijen, and
  Swamy]{HicksBGLS14}
M.~Hicks, G.~M. Bierman, N.~Guts, D.~Leijen, and N.~Swamy.
\newblock Polymonadic programming.
\newblock In \emph{MSFP}, 2014.

\bibitem[Katsumata(2014)]{Katsumata14}
S.~Katsumata.
\newblock Parametric effect monads and semantics of effect systems.
\newblock In \emph{POPL}, 2014.

\bibitem[Lunn et~al.(2000)Lunn, Thomas, Best, and Spiegelhalter]{LunnTBS00}
D.~J. Lunn, A.~Thomas, N.~Best, and D.~Spiegelhalter.
\newblock Win{BUGS} - {A} bayesian modelling framework: Concepts, structure,
  and extensibility.
\newblock \emph{Statistics and Computing}, 2000.

\bibitem[McSherry(2009)]{PINQ09}
F.~McSherry.
\newblock Privacy integrated queries: an extensible platform for
  privacy-preserving data analysis.
\newblock In \emph{International Conference on Management of Data}, 2009.

\bibitem[McSherry and Talwar(2007)]{MT07}
F.~McSherry and K.~Talwar.
\newblock Mechanism design via differential privacy.
\newblock In \emph{FOCS}, 2007.

\bibitem[Minka et~al.(2012)Minka, Winn, Guiver, and Knowles]{InferNET12}
T.~Minka, J.~Winn, J.~Guiver, and D.~Knowles.
\newblock {Infer.NET 2.5}, 2012.
\newblock URL \url{http://research.microsoft.com/infernet}.
\newblock MSR.

\bibitem[Pfeffer(2001)]{Pfeffer01}
A.~Pfeffer.
\newblock {IBAL:} {A} {P}robabilistic {R}ational {P}rogramming {L}anguage.
\newblock In \emph{IJCAI}, 2001.

\bibitem[Reed and Pierce(2010)]{ReedPierce10}
J.~Reed and B.~C. Pierce.
\newblock Distance {M}akes the {T}ypes {G}row {S}tronger: {A} {C}alculus for
  {D}ifferential {P}rivacy.
\newblock In \emph{ICFP}, 2010.

\bibitem[Sato()]{sato2016}
T.~Sato.
\newblock Approximate {R}elational {H}oare {L}ogic for {C}ontinuous {R}andom
  {S}amplings.
\newblock \emph{CoRR}, abs/1603.01445.

\bibitem[Staton et~al.(2016)Staton, Yang, Heunen, Kammar, and Wood]{lics16}
S.~Staton, H.~Yang, C.~Heunen, O.~Kammar, and F.~Wood.
\newblock Semantics for probabilistic programming: higher-order functions,
  continuous distributions, and soft constraints.
\newblock In \emph{LICS}, 2016.

\bibitem[Tolpin et~al.(2015)Tolpin, van~de Meent, and Wood]{TolpinMW15}
D.~Tolpin, J.~van~de Meent, and F.~Wood.
\newblock Probabilistic {P}rogramming in {A}nglican.
\newblock In \emph{{ECML} {PKDD}}, 2015.

\bibitem[Toronto et~al.(2015)Toronto, McCarthy, and Horn]{TorontoMH15}
N.~Toronto, J.~McCarthy, and D.~V. Horn.
\newblock Running {P}robabilistic {P}rograms {B}ackwards.
\newblock In \emph{ESOP}, 2015.

\bibitem[Williams and McSherry(2010)]{DBLP:conf/nips/WilliamsM10}
O.~Williams and F.~McSherry.
\newblock Probabilistic {I}nference and {D}ifferential {P}rivacy.
\newblock In \emph{NIPS}, 2010.

\bibitem[Zhang et~al.(2014)Zhang, Cormode, Procopiuc, Srivastava, and
  Xiao]{Zhang:2014:PrivBayes}
J.~Zhang, G.~Cormode, C.~M. Procopiuc, D.~Srivastava, and X.~Xiao.
\newblock Priv{B}ayes: Private data release via bayesian networks.
\newblock In \emph{SIGMOD}, 2014.

\bibitem[Zhang et~al.(2016)Zhang, Rubinstein, and Dimitrakakis]{ZhangRD16}
Z.~Zhang, B.~I.~P. Rubinstein, and C.~Dimitrakakis.
\newblock On the {D}ifferential {P}rivacy of {B}ayesian {I}nference.
\newblock In \emph{AAAI}, 2016.

\bibitem[Zheng(2015)]{Zheng16}
S.~Zheng.
\newblock The differential privacy of {B}ayesian inference, 2015.
\newblock URL \url{http://nrs.harvard.edu/urn-3:HUL.InstRepos:14398533}.
\newblock Bachelor's thesis, Harvard College.

\end{thebibliography}
\end{document}